\documentclass[sigplan,10pt]{acmart}
\renewcommand\footnotetextcopyrightpermission[1]{}

\newcommand{\oursys}{{Rubick}}

\newcommand{\eg}{\textit{e.g.}}
\newcommand{\ie}{\textit{i.e.}}

\newcommand{\ourSys}{{\textit{\oursys}}}

\newcommand{\faye}[1]{\textcolor{black}{#1}}
\newcommand{\hanyu}[1]{\textcolor{black}{#1}}
\newcommand{\revision}[1] {\textcolor{black}{#1}}
\newcommand{\newrevision}[1] {\textcolor{black}{#1}}
\newcommand{\xinyi}[1] {\textcolor{black}{#1}}

\renewcommand{\paragraph}[1]{{\bf \noindent #1 \hspace{5pt}}}

\newenvironment{packed_itemize}
{
  \begin{list}{\labelitemi}{\leftmargin=1.0em}
    \setlength{\itemsep}{4pt}
    \setlength{\parskip}{0pt}
    \setlength{\parsep}{0pt}
    \setlength{\headsep}{0pt}
    \setlength{\topskip}{0pt}
    \setlength{\topsep}{0pt}
    \setlength{\partopsep}{0pt}
  }{\end{list}}

\usepackage[normalem]{ulem}
\usepackage{amsmath,mathtools}

\usepackage{amssymb}
\usepackage{multirow,booktabs}
\usepackage[linesnumbered,ruled,vlined]{algorithm2e}
\usepackage{float}
\usepackage{graphicx}
\usepackage{subfig}
\usepackage{mathrsfs}
\usepackage{pifont}
\usepackage{booktabs} 
\usepackage{makecell}
\begin{document}

\title{\ourSys{}: Exploiting Job Reconfigurability for\\Deep Learning Cluster Scheduling}
\author{
    {\rm Xinyi Zhang$^{\dag}$$^{*}$, Hanyu Zhao$^{\star}$, Wencong Xiao$^{\star}$, Xianyan Jia$^{\star}$\\Fei Xu$^{\dag}$, Yong Li$^{\star}$, Wei Lin$^{\star}$, Fangming Liu$^{\ddag}$} \\ 
    \vspace{1em}
    $^\dag$East China Normal University, $^\star$Alibaba Group\\$^\ddag$Huazhong University of Science and Technology, and Peng Cheng Laboratory
    \vspace{2em}
}



\begin{abstract}

\hanyu{The era of large deep learning models has given rise to advanced training strategies such as 3D parallelism and the ZeRO series. These strategies enable various (re-)configurable execution plans for a training job,}
which exhibit \faye{remarkably different requirements of multiple resource types}. Existing cluster scheduling systems, however, treat such reconfigurable training jobs as black boxes: they rely on users to choose execution plans statically, and then make resource allocations without awareness of the chosen plans and their resource requirements. This approach results in mismatches between execution plans and resources, making both training performance and cluster utilization far from optimal.

We introduce \ourSys{}, a cluster scheduling system for deep learning training that exploits the reconfigurability to improve job performance and cluster efficiency. \ourSys{} incorporates the job execution planning as a new dimension in cluster scheduling, by continuously reconfiguring jobs’ execution plans and tuning multi-resource allocations across jobs jointly.
Such a co-optimization is navigated by a performance model that understands the diverse resource requirements and performance characteristics of different jobs and execution plans.
\ourSys{} exploits such a model to make performance-aware scheduling decisions to maximize cluster throughput while providing performance guarantees to individual jobs.
Evaluations on a $64$-GPU high-performance training cluster 
show that \ourSys{} improves average job completion time and makespan by up to $3.2\times$ and $1.4\times$, respectively, compared against state-of-the-art systems.

\end{abstract}

\settopmatter{printfolios=true,printacmref=false}
\maketitle
\begingroup
\renewcommand\thefootnote{\fnsymbol{footnote}}
\footnotetext[1]{Work done during internship at Alibaba Group.}
\endgroup
\pagestyle{plain}

\section{Introduction}
\label{sec:intro}



\revision{With the dominance of Transformer architectures~\cite{vaswani2017attention} in terms of model performance across a variety of applications, deep learning (DL) has recently entered an era characterized by exponentially increasing model sizes, which further escalates training resource (\eg, GPU) requirements~\cite{Radford2019LanguageMAGPT2, swintransformer2021, bert19}.} 
\revision{To facilitate efficient large-scale DL training, organizations such as  Microsoft~\cite{jeon2019analysis} and Alibaba~\cite{weng2022mlaas} have built multi-tenant shared GPU clusters, thereby improving resource utilization.}


\revision{Numerous research efforts have been devoted to optimizing job \emph{execution plans} for large model training.}
For instance, several studies concentrate on scheduling and partitioning operators and tensors to attain better performance~\cite{zheng2022alpa, unger2022unity, jia2022whale}, while others focus on optimizing GPU memory usage by eliminating duplicate states~\cite{rajbhandari2020zero}, recomputing activation~\cite{chen2016training}, and offloading~\cite{ren2021zero,rajbhandari2021zero}.
These cutting-edge techniques have proven to be effective in improving the performance of DL jobs on \revision{dedicated} resources. 
However, they fall short in dynamic shared clusters, where resource availability can vary significantly during job training~\cite{weng2022mlaas}.
\revision{This is mainly because the training paradigm follows a \emph{compile-and-run} approach. Specifically, an execution plan is pre-compiled at job launch time and then runs iteratively until completion on fixed allocated resources. Such an approach fundamentally impedes the possibility of exploiting resource dynamics efficiently.}

\begin{figure}[tb]
  \centering
  \includegraphics[width=1\linewidth]{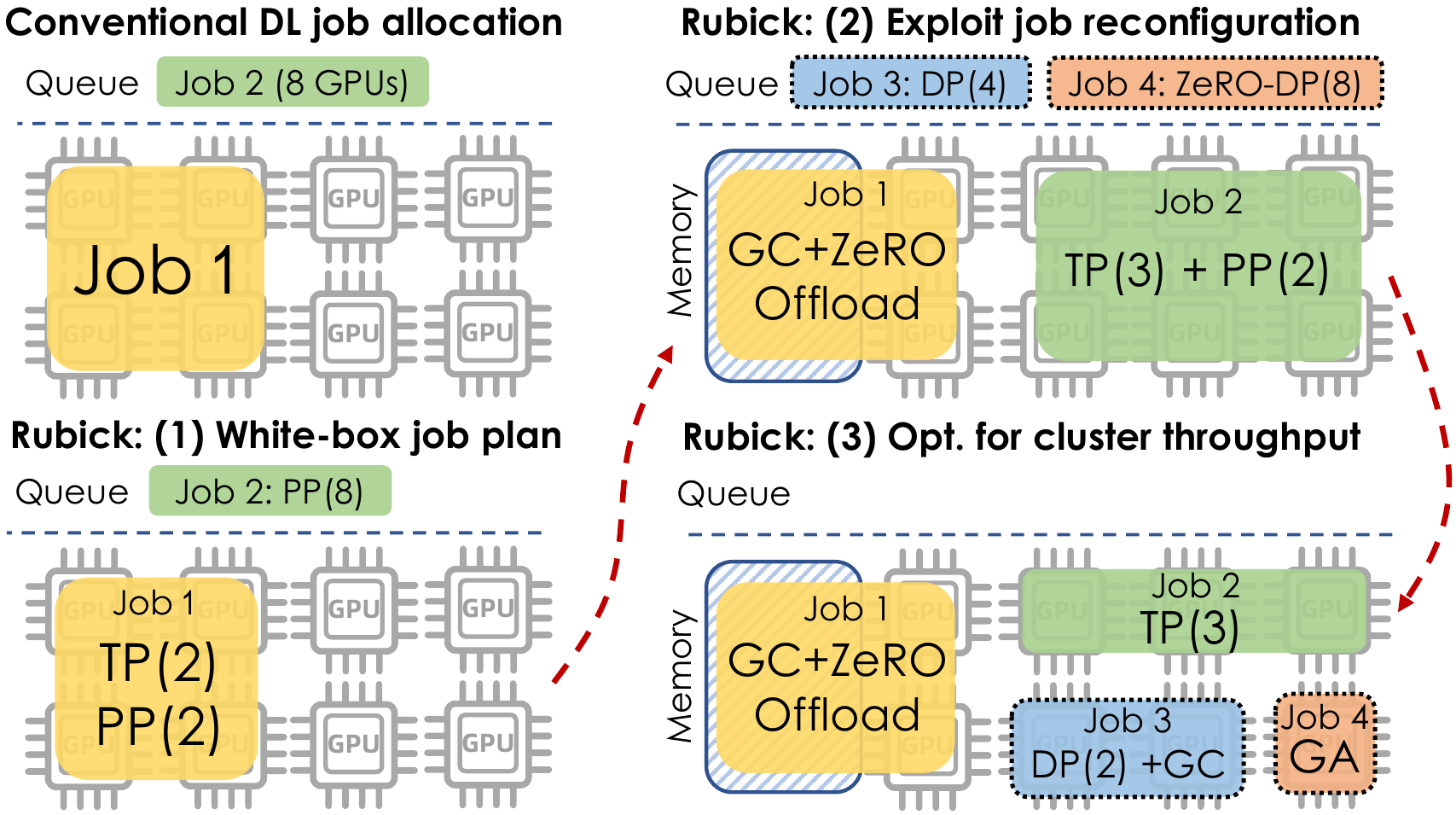}
  \caption{Overview of \ourSys{}. Its fundamental capability lies in leveraging white-box execution plans to enable job reconfiguration and cluster-level throughput optimization. \newrevision{Job execution plans (\eg, TP, PP, GC) are elaborated in Sec.~\ref{sec:motivation-background}.}}
  \label{fig:arch}
\end{figure}


\revision{From a cluster management standpoint, DL training jobs typically request a predetermined amount of resources and must wait the availability of all resources due to the gang-scheduling requirement~\cite{jeon2019analysis, weng2022mlaas}. To reduce the job queuing delay, several recent studies} have proposed elastic resource scheduling for distributed data-parallel jobs~\cite{hwang2021coddl, qiao2021pollux, li2023easyscale}. However, the cluster scheduler only scales the number of training workers without considering the execution plan, resulting in two constraints arising from the \revision{DL} resource characteristics implicit in the job execution plans. 
\emph{First,} different execution plans come with diverse resource requirements. 
Despite the primary concern about the number of GPUs, these plans also impact the multi-resource requirements (\ie, GPU, CPU, memory, network).
For example, switching from tensor parallelism~\cite{megatron2019} to ZeRO-Offload~\cite{ren2021zero} effectively reduces the demand for GPUs, \newrevision{but incurs} higher memory consumption in exchange. 
\emph{Second,} there is no single execution plan that can be considered as optimal under all GPU resource allocations. 
As evidenced by Fig.~\ref{subFig:moti_bert}, ZeRO-DP~\cite{rajbhandari2020zero} is the best plan given eight GPUs when training the GPT-2 model, while tensor model parallelism and ZeRO-Offload are the best for four GPUs and one GPU, respectively.
Such two observations above imply an interesting interplay between a training job's execution plan and the resource allocation to it. Specifically, given a limited amount of available resources, it is possible to \emph{adapt the execution plan to the resource} by choosing a plan whose multi-resource demand matches the available resources the best.
On the other hand, when resources are abundant, it is also possible to \emph{adapt the resource allocation to the plan} by choosing a plan that exhibits the best performance, and then allocating resources according to the demand of that plan.

Unfortunately, \revision{such an opportunity above} is largely overlooked in current DL training clusters, where the decisions for execution plans and resource allocations are made separately.
The execution plans are chosen by users statically, without the knowledge about the dynamics of cluster resources, prohibiting \emph{reconfiguration} of the plan from adapting to the resources. Meanwhile, the resource allocations are either following user-specified requirements or tuned by cluster schedulers. Despite knowing the plans they choose, users typically do not have the knowledge or profiling expertise to understand the resource demands of the plans. Current cluster schedulers, on the other hand, even have no information about job's execution plans. Either way, it is also difficult to optimize the resource allocations according to the execution plans.

We introduce \ourSys{}, a novel cluster scheduling system that exploits the reconfigurability of DL training to bridge the gap between intra-job execution planning and inter-job resource scheduling.
As illustrated in Fig.~\ref{fig:arch}, unlike conventional schedulers that treat DL jobs as \emph{pre-defined static} execution plans for scheduling, \ourSys{} performs a \emph{white-box} approach to co-optimize cluster resources and training strategies of jobs dynamically through execution plan reconfiguration.
Such a design enables \ourSys{} to continuously reconfigure the execution plans for individual jobs and reallocate multi-dimensional resources across jobs co-adaptively.

To help \ourSys{} understand the multi-resource demands of various execution plans, we establish a resource-performance model for a series of widely-used training strategies \faye{to characterize} their fine-grained behaviors carefully. With such a model, \ourSys{} predicts the performance of each job with any combinations of the execution plan and resource allocation.
Guided by such performance predictions, \ourSys{} further employs a \revision{performance-aware} scheduling policy to search for optimized execution plans efficiently for each job while adjusting the multi-resource allocations across jobs, with the
aim of maximizing cluster throughput while guaranteeing the
service level agreement (SLA) to individual jobs.


\revision{We evaluate \ourSys{} on a 64-GPU cluster to show the advantages of the reconfiguration ability
and job-plan-aware scheduling policy with micro-benchmarks and real workloads. 
Trace evaluations show that \ourSys{} preserves the SLA guarantees for jobs and improves the average job completion time
(JCT) by up to $3.2\times$ compared to state-of-the-art DL cluster schedulers (\hanyu{Sia~\cite{subramanya2023sia}}, Synergy~\cite{synergy22}, and AntMan~\cite{antman2020}).}

The contributions of this paper are summarized as follows.
\begin{packed_itemize}
    \item We reveal the diverse multi-resource requirements of various training strategies and identify the interplay between execution plans and resource allocations for DL training.
    \item We propose a system architecture to embrace job plan reconfiguration as a new dimension in cluster scheduling.
    \item We design a performance model and a scheduling policy to maximize job performance and cluster throughput by co-optimizing execution plans and resource allocations.
    \item We implement and evaluate \ourSys{} to show its advantages over reconfigurability-agnostic systems. 
\end{packed_itemize}

\section{Background and Motivation}
\label{sec:motivation}



\subsection{\hanyu{Large Model Training in GPU Clusters}}
\label{sec:motivation-background}

\begin{figure*}[t]
  \begin{minipage}{0.29\textwidth} %
    \centering 
    \hspace{0.6in}
    \includegraphics[width=1\textwidth]{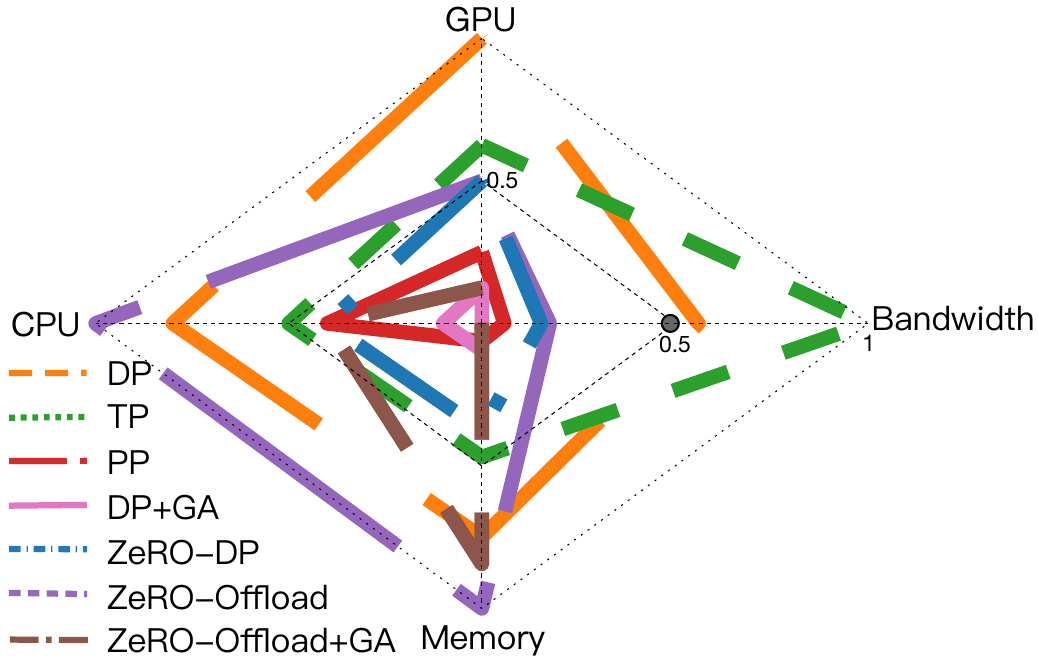} 
    \caption{Consumption of each resource type for GPT-2 using various training execution plans, normalized to the highest value ($8$ GPUs, $10$ CPUs, $3.2$ GB memory, and $30$ GB/s bandwidth).}
    \label{fig:moti_demand}
  \end{minipage} \hspace{+6pt}
  \begin{minipage}{0.68\textwidth}
    \centering
    \subfloat[RoBERTa]{
    \label{subFig:moti_roberta}
    \includegraphics[width=0.45\textwidth]{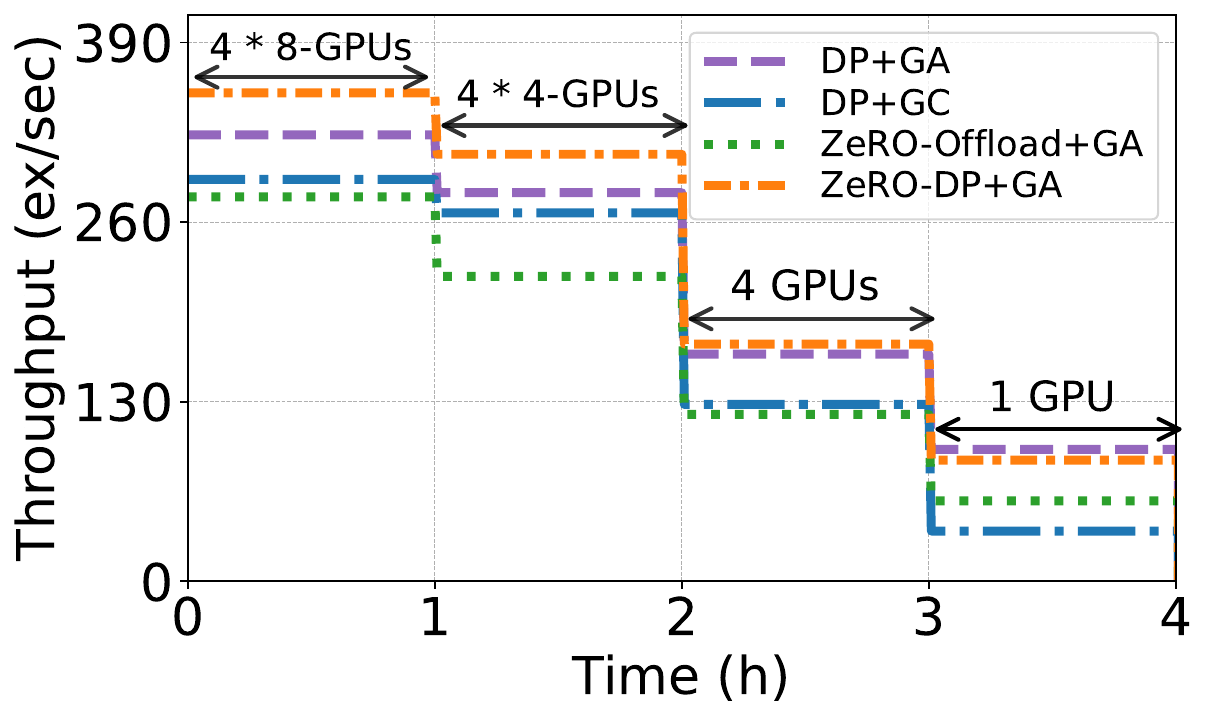} 
    }
    \subfloat[T5]{\label{subFig:moti_bert}
    \includegraphics[width=0.5\textwidth]{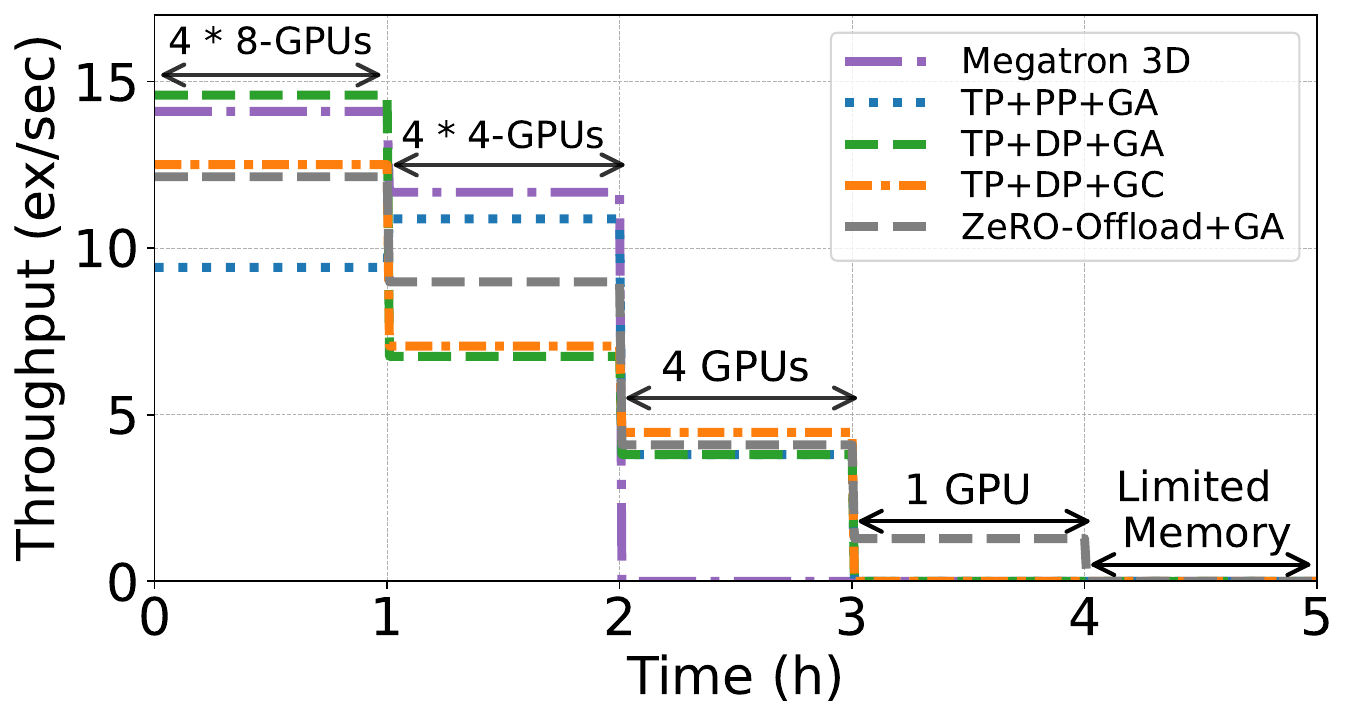}
    }
    \caption{\xinyi{Throughput variation using various execution plans with changing resource limits. The first hour is using $4$ servers with $8$ A800 GPUs \newrevision{for each}, and the second hour is using $4$ servers with $4$ A800 GPUs. The rest are using a $4$-A800 server. TP+DP/PP means using TP inside nodes and DP/PP across nodes. Megatron 3D adopts a feasible TP+PP configuration such that each partition fits in a GPU, then scaling out using DP.}}\label{fig:moti_throughput}
  \end{minipage}%
\end{figure*}


DL training often involves millions of iterations, and each is called a \textit{mini-batch}. 
Typically, a basic mini-batch contains three phases. 
Firstly, current model scores are calculated by training samples and model weights using a DAG 
of operators, known as a \textit{forward pass}. 
Secondly, a loss error between the current score and the objective is produced to spread backwards through the model to generate gradients, called a \textit{backward pass}. 
Finally, the model parameters are updated by the gradients scaled by an \textit{optimizer}.

To scale the model training to distributed GPUs, 
\textit{data parallelism} (DP) uses multiple workers each executing the full model with a subset of a mini-batch, and
synchronizes gradients across workers per mini-batch after the backward pass~\cite{pytorchddp2020},
which, however, introduces significant cost on network traffic and GPU memory for large models.
Therefore, \hanyu{more advanced training strategies are proposed to further scale the training to even larger models}. 
The current state-of-the-art strategy in large model training is \textit{3D parallelism}, 
which combines \textit{tensor model parallelism} (TP)~\cite{megatron2019,tofu2019} and \textit{pipeline parallelism} (PP)~\cite{narayanan2019pipedream,huang2019gpipe} with \newrevision{DP}.
Tensor model parallelism partitions the computation of a specific operator, such as matrix multiplication, in non-batch axes, to multiple GPUs.
Pipeline parallelism groups model operators into stages and places them on different GPUs. It then splits a mini-batch into a number of \textit{micro-batches} for forward-backward computation across GPUs.
\hanyu{The degrees of the three parallelism dimensions are typically referred to as \emph{DP/TP/PP sizes}; \ie, number of model replicas, number of model partitions, and number of pipeline stages, respectively.}
The combinations of such strategies are either specified by users~\cite{megatron2019, jeff2020deepspeed} or automated~\cite{zheng2022alpa, unger2022unity, jia2022whale}
to efficiently scale the training on trillions of parameters over hundreds or thousands of GPUs.

\revision{Another thread of techniques focuses on saving GPU memory consumption in \hanyu{large model} training.}
Gradient accumulation (GA)~\cite{keskar2017largebatch} divides a mini-batch into micro-batches and aggregates per-worker gradients locally before global synchronization.
Gradient checkpointing (GC)~\cite{chen2016training} chooses to preserve a subset, instead of all, of the intermediate results in forward passes (called activations) to reduce GPU memory.
It recomputes the missing activations on-demand in backward passes. GA and GC can be used in conjunction with the parallel strategies presented before.
ZeRO-DP~\cite{rajbhandari2020zero} deduplicates the redundant states (\ie, optimizer states, gradients, and weight parameters) of DP by slicing them across all GPUs\footnote{There are several ZeRO-DP variants, and we refer to ZeRO-2 by default.}.
Going further, ZeRO-Offload~\cite{ren2021zero} keeps only the forward-backward pass in GPU, offloads the activations and states to host memory, and updates the parameters using CPUs.

\revision{
To submit a job to a GPU cluster, users need to specify the required multi-dimensional resources for a worker, and the number of workers for distributed jobs~\cite{weng2022mlaas, synergy22}.
For instance, a typical distributed training job can request $2$ workers, each with $8$ GPUs, $16$ CPUs, and $100$ GB memory.
Cluster scheduler launches jobs at the availability of all resources~\cite{jeon2019analysis}. For DP jobs, the support of GPU training elasticity~\cite{qiao2021pollux, li2023easyscale} has been explored, by scaling the number of training workers during the execution of jobs.  
}

\subsection{Opportunity and Challenge}
\label{sec:motivation-opportunity}

\paragraph{Opportunity: diverse multi-resource demands of different execution plans.}
The application of the training strategies above can produce diverse execution plans for model training. 
A notable variance exists in the resource types and quantities required for these plans. 
Fig.~\ref{fig:moti_demand} shows \revision{the resource consumption of multiple execution plans for training a GPT-2 model with the minimum A800 GPUs with a global batch size of 16.}
For example, ZeRO-Offload uses the most CPU and memory resources for parameter updates and model states offloading.
In comparison, under nearly the same number of GPUs, TP uses more bandwidth for heavier communication volumes, while using only half of the CPUs and memory.

Despite the diverse resource demands of execution plans, there exist significant gaps between the execution planning of training jobs and the resource allocation in shared GPU clusters.
From the perspective of cluster schedulers, they perceive DL training jobs as black-box tasks with fixed resource requirements, 
disregarding the variability in resource demands of various execution plans. 
On the other hand, the job's standpoint involves manual or automatic decisions of the execution plan before training. 
This approach assumes that the cluster is dedicated and exclusive, which does not hold in shared clusters where resource supply is typically dynamic and unknown to users~\cite{weng2022mlaas,weng2023fgd}.
Consequently, DL job execution is often suboptimal. 
When resources are limited,
a job may be delayed due to an excess of requested resources, or run with degraded performance due to the mismatch between the resources and the requirement of its execution plan.
Conversely, when resources are overabundant, jobs may not fully utilize such resources due to the fixed resource request or the inefficiency of the execution plan. 

This presents great opportunities for cluster schedulers to leverage the reconfiguration capabilities of DL jobs to narrow the gaps. 
By doing so, jobs can adapt to the dynamic multi-resource availability with efficient training strategies and optimizations properly, while DL cluster schedulers could view the job execution plan and resource demand transparently to optimize scheduling policies, thereby improving cluster efficiency and expediting job completion.

\paragraph{Challenge: complex performance characteristics of model-plan-resource combinations.}
We conduct two motivating experiments towards a deeper understanding of the performance characteristics of different models and execution plans.
\xinyi{
We first train a RoBERTa model with multiple plans, and change the limit of a certain resource type in each stage. Fig.~\ref{subFig:moti_roberta} shows that the performance of the plans and their relative rankings vary across stages. In the first three stages, where the GPUs are sufficient with abundant bandwidth, the best plans are ZeRO-DP due to its reduction in the optimizer time, which scales favorably with the increased number of GPUs for the model state partitioning. With GPUs reduced to $1$ in the fourth stage, ZeRO-DP performs worse with increased optimizer time, making DP+GA the new best.}

\xinyi{Fig.~\ref{subFig:moti_bert} shows the same process with a larger model, T5, for comparison. In the initial two stages, where the GPUs are distributed, the best plans are 3D parallelism with different DP/TP/PP sizes. This is because the performance is constrained by the bandwidth limits across nodes, thus depending on the communication volume under different 3D parallelism configurations. With a single server in the third stage, TP+DP+GC becomes the best plan with its modest recomputation overhead when the GPU memory is limited. With GPU reduced to $1$ in the fourth stage, ZeRO-Offload becomes the only plan that can still continue the training with the use of CPUs and memory. In the final stage, we further limit the memory to $10$ GB, which makes ZeRO-Offload fail.}


\xinyi{We also observe that the execution plans exhibit different performance characteristics with a different model. 
For example, ZeRO-Offload nearly always performs the worst on RoBERTa, while this is not the case for T5. 
Moreover, the two models show different sensitivity to different execution plans. The max performance gap between plans in the same stage is up to $1.7\times$ \emph{vs.} $2.7\times$ for T5 and RoBERTa, respectively, showing different levels of benefits from reconfiguration.}

\paragraph{\revision{Summary.}}
The observations above show the complex performance characteristics of different combinations of models, plans, and resources: each single job can have varying best plans with changing resource availability; moreover, different jobs also exhibit different sensitivity to changing resource and execution plans.
Such complexity is determined inherently by the heterogeneous model structures, diverse training behaviors of the plans, and their different resource usage patterns.
The scheduler needs to understand such performance characteristics to derive high-quality resource allocations and execution plans. However,
it is nearly impossible to enumerate every combination to
measure the real performance, considering the intractable search space of models, plans, and especially the multiple types of resources in a large cluster. 
This motivates a performance-modeling approach
to predict the performance of various plans and resources for a job with 
a limited number of sampled configurations for measurement.

\section{System Overview}
\label{sec:overview}

\begin{figure}[t]
  \centering
  \includegraphics[width=0.85\linewidth]{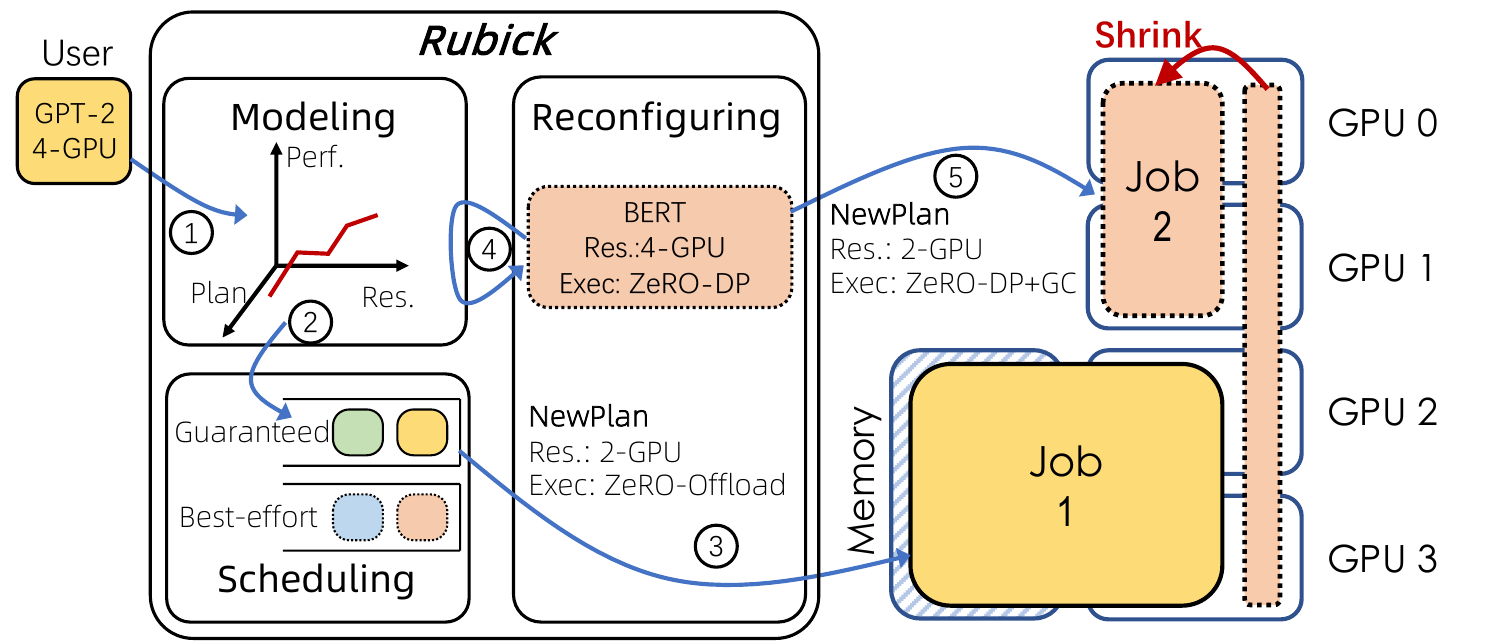}
  \caption{\ourSys{} architecture and scheduling workflow.}
  \label{fig:systemArchitecture}
\end{figure}

Going beyond the traditional responsibility of allocating resources to incoming jobs,
\ourSys{} also takes control of job execution planning. It continuously adjusts the resource allocation and reconfigures the execution plans jointly for all running jobs.
As shown in Fig. \ref{fig:systemArchitecture}, \ourSys{} operates in three main phases: \revision{First,} profiling and performance modeling for new model types (\ding{192}); \revision{second,} allocating resources and choosing execution plans for each job with a scheduling policy (\ding{193}); and finally, launching new jobs (\ding{194}) or reconfiguring running jobs (\ding{195}) per the scheduling decision.


\ourSys{} supports a series of widely-used execution plans including (1) Megatron-style 3D parallelism (DP-TP-PP)~\cite{megatron2019,megatron2021}, (2) ZeRO-DP~\cite{rajbhandari2020zero} and ZeRO-Offload~\cite{ren2021zero} based on DP, (3) gradient accumulation~\cite{keskar2017largebatch} or checkpointing~\cite{chen2016training} (GA/GC) based on DP or the ZeRO-series (\eg, ZeRO-DP plus GA).
\ourSys{} can reconfigure a job by switching among different types of execution plans; for 3D parallelism, in particular, \ourSys{} also supports changing the DP/TP/PP size parameters.
\ourSys{} keeps the global batch size of a job unchanged during reconfiguration, thus not affecting the training convergence.


\ourSys{} establishes a performance model for reconfigurable DL training (Sec.~\ref{sec:model}) to enable performance-aware scheduling. The model captures the fine-grained behaviors of the various training strategies, and also the impact of resource variations on their performance. \faye{It needs to be fitted for each DL model using a few sampled performance points under several configurations (\ie, execution plans and resources). 
Once fitted, the model can predict the performance with other unseen configurations. 
In particular, the model can be reused throughout the lifetime of a job with continuous reconfigurations. It can also be reused across multiple jobs of the same model type,} \ie, jobs with exactly the same model architecture but possibly different hyper-parameters like learning rate.
\ourSys{} allows users to associate such jobs with a model-type flag to facilitate such reuse.


Leveraging the performance model, the \ourSys{} scheduler further allocates resources and reconfigures execution plans for the jobs (Sec.~\ref{sec:scheduler}). 
\ourSys{} redefines the design goals of conventional schedulers \revision{by taking} into account the new scheduling dimension of execution planning.
\ourSys{} provides SLA guarantees to jobs, during continuous reconfigurations, by ensuring that their performance would be no worse than that with a baseline resource amount and execution plan specified by users.
That is, \ourSys{} can deliver the same or better performance with even \emph{fewer} resources by \faye{identifying} better execution plans.
With such guarantees for individual jobs,
\ourSys{} further co-optimizes the resource allocation and the execution planning for each job to \faye{maximize cluster throughput,} 
\ie, the aggregated performance of all jobs.

\section{Modeling Reconfigurable DL Training}
\label{sec:model}


In this section, we model and predict the training throughput using different combinations of strategies and resource allocations. 
We denote each fittable parameter in our model as $k$ plus a certain subscript to distinguish it from other model- or environment-related constants (summarized in Table~\ref{tab:model_sum}).

\begin{figure}[t]
  \centering
  \includegraphics[width=0.4\textwidth]{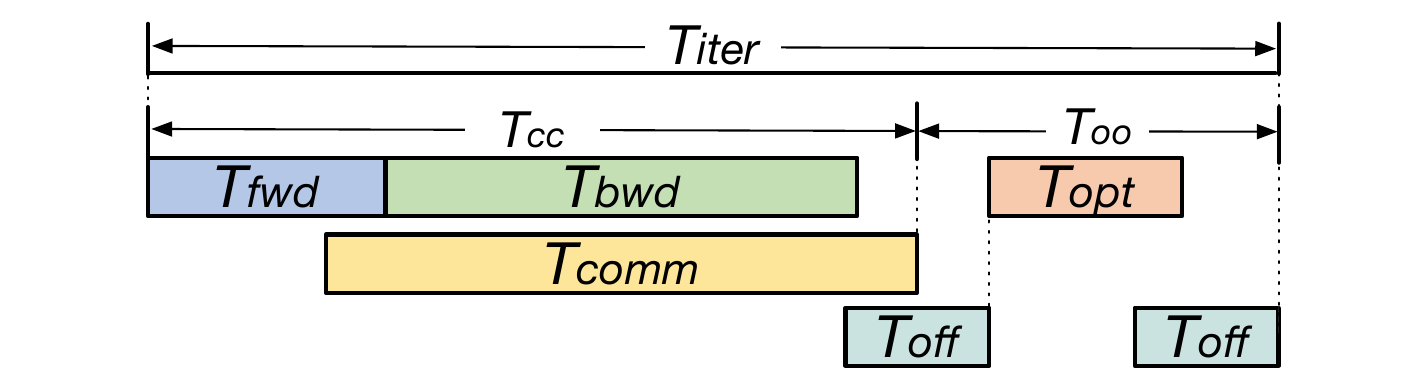}
  \caption{ Simplified illustration of the performance model.
Note that the overlapping of the parts only means the overlapping of their time spans; the real execution is not necessarily
overlapped, which depends on the specific strategy.}\label{fig:model-parts}
\end{figure}

We aim to predict the time spent per training iteration, $T_{iter}$, and then calculate the throughput as $\mathrm{THROUGHPUT}=b/T_{iter}$,
where $b$ is the global batch size.
$T_{iter}$ is generally comprised of the following parts: $T_{fwd}$, the time for forward pass computation; $T_{bwd}$, backward pass computation; $T_{comm}$, network communication; $T_{opt}$, optimizer; and $T_{off}$, model states offloading.
$T_{iter}$ is typically not a sum of these parts, because they are usually overlapped with each other. Fig. \ref{fig:model-parts} illustrates a simplified example of the combination of these parts to give readers an intuition.
We will first model each of these parts, and then show how to combine them into $T_{iter}$.

\subsection{Modeling Computation and Communication}
\label{sec:model-computate}

\paragraph{Modeling $T_{fwd}$.} 
The time for forward pass $T_{fwd}$ under 3D parallelism can generally be obtained from profilers provided by DL frameworks (\eg, in DeepSpeed) on a node with a given global batch size. We scale up or down $T_{fwd}$ linearly to the actual per-GPU batch size \revision{for data parallelism, and to per-GPU tensor shard size for tensor parallelism}. Besides, we have special treatments for the following two strategies.

\emph{Pipeline Parallelism.} \revision{PP averagely places the layers among GPUs.}  When \revision{profiling with $g_p$ GPUs}, the forward time \revision{provided} by the framework (denoted as $t_{p}$) is usually the time \revision{for a single GPU to process a micro-batch, with $l/g_p$ layers placed on it, where $l$ is the total number of layers.} The complete $T_{fwd}$ for PP includes the time taken for the first micro-batch to be processed sequentially on each GPU, and that for all GPUs to serially process the other micro-batches~\cite{narayanan2019pipedream}.  \revision{Besides, $T_{fwd}$ is linear to the per-GPU number of layers. We then have $T_{fwd}=t_{p}\cdot g_p /p \cdot (m+p-1)$}, where $m$ and $p$ are the numbers of micro-batches and PP size, respectively.

\emph{Gradient Accumulation.}  GA aggregates per-GPU gradients over multiple passes. Therefore, the total forward time is $T_{fwd} \cdot a$, where $a$ is the number of accumulation steps.

\paragraph{Modeling $T_{bwd}$.}
$T_{bwd}$ is the time for computing the gradients during the backward pass.
Transformer-based models are primarily comprised of matrix multiplication operations, where the time required for gradient computation can be generally considered to be proportional to $T_{fwd}$, \ie, $T_{bwd}= k_{bwd}\cdot T_{fwd}$.
An exception is gradient checkpointing (GC): GC recomputes activations during the backward pass. The time cost for the extra computation is typically equal to the time $T_{fwd}$~\cite{chen2016training}. Therefore, when GC is used, modeling the $T_{bwd}$ requires adding the time required for a forward pass.


\paragraph{Modeling $T_{comm}$.}
\hanyu{The communication time $T_{comm}$ involves those for data, tensor, and pipeline parallelisms. For each part, $T_{comm}$ is in general estimated as $T_{comm}=V/B$,}
where ${V}$ is the volume of the data to transfer \hanyu{between each pair of GPUs and ${B}$ is the corresponding bandwidth.}

We discuss how to model $B$ first. \hanyu{For each type of communication (DP/TP/PP),}
we basically use the bottleneck bandwidth of the GPUs \hanyu{involved in the communication}, \ie, the lowest bandwidth among all pairs of GPUs.
For example, when all GPUs are co-located on the same node,
the data can be transferred via a high-speed connection like NVLink.
In this case, we use the intra-node bandwidth $B_{intra}$ as $B$.
However, when the GPUs are spread on multiple nodes, the communication is largely dominated by the bandwidth between nodes because the speed is much slower than NVLink. Hence we use inter-node bandwidth $B_{inter}$ here. 
\hanyu{Note that different types of communication may use different $B$ values. For example, TP is typically restricted inside each node while PP can be distributed across nodes~\cite{megatron2021}. In this case, TP and PP will use $B_{intra}$ and $B_{inter}$, respectively.}
The values of $B_{intra}$ and $B_{inter}$ are measured on the cluster offline. 


Next, we model the communication volume $V$ for different strategies respectively.
\hanyu{When the parallelism size of any dimension is $1$}, then the corresponding $V$ is $0$.

\emph{Data Parallelism.}  
DP typically uses the ring AllReduce algorithm to synchronize the gradients, where each model replica sends and receives $2(d-1)/d$  times gradients ($d$ being the DP size). The gradients generated during the entire backward pass are approximately as large as the parameter size. \hanyu{Considering that the gradients are partitioned and synchronized in parallel across TP and PP partitions, we have $V_{dp}=P \cdot 2(d-1)/(d \cdot t \cdot p)$, where $P$ is the total parameter size, and $t$ and $p$ are TP and PP sizes, respectively. This rule also applies to the ZeRO series as they are based on DP.}




\begin{table}[t]
\renewcommand{\arraystretch}{1.2} 
\centering
\caption{Summary of performance model parameters.}
\footnotesize
\begin{tabular}{cl|l}
\toprule[1pt]
\multicolumn{2}{c|}{Fittable}                                  & $k_{bwd}, k_{sync}, k_{opt}, k_{opt\_off}, k_{off}, k_{swap}, k_{const}$               \\ \hline
\multicolumn{1}{c|}{\multirow{4}{*}{Job}} & Model    & $s$ (seq), $h$ (hidden), $l$ (layers), $P$ (param size)          \\ \cline{2-3} 
\multicolumn{1}{c|}{}                               & Resources & $g$ (GPU), $c$ (CPU)  \\ \cline{2-3} 
\multicolumn{1}{c|}{}                               & Parallelism & $d, t, p$ (3D-parallel sizes, \hanyu{$d \cdot t \cdot p = g$}) \\ \cline{2-3} 
\multicolumn{1}{c|}{}                               & Others & $b$ (batch size), $m$ (micro-batch num), $a$ (GA steps)  \\ \hline
\multicolumn{2}{c|}{Environment}                       & $B_{intra},B_{inter},B_{pcie}$          \\ 
\bottomrule[1pt]
\end{tabular}
\label{tab:model_sum}
\end{table}

\emph{Tensor Parallelism.} The communication volume for TP depends on the size of output tensor \hanyu{of a transformer layer,} which is $b \cdot s \cdot h$~\cite{vaswani2017attention} \hanyu{when not sliced}, where $b$, $h$, and $s$ represent the batch size, hidden size, and sequence length, respectively. \hanyu{Each layer involves in total $4$ communication operations in the forward and backward passes~\cite{megatron2019}. Considering the output tensor and the batch are partitioned by TP and DP, respectively,
we have $V_{tp}=4 \cdot 2 \cdot (t-1) \cdot b \cdot s\cdot h \cdot l / (d \cdot t)$ (this volume is not divided by the PP size $p$ because the TP communications across pipeline stages are serialized).}

\emph{Pipeline Parallelism.}
Micro-batches need to wait for the communication from other pipeline stages after finishing the forward/backward pass for the current micro-batch.
The communication volume for each micro-batch between each consecutive pair of devices is $b/m \cdot s \cdot h$. \hanyu{PP communication is involved in both forward and backward passes, and the tensors are partitioned by DP and TP along the batch size and operator dimensions.}
We thus have $V_{pp}= 2 \cdot p \cdot b \cdot s \cdot h / (d \cdot t)$\footnote{We model the commonly used 1F1B strategy for PP~\cite{narayanan2019pipedream}. This formula only considers the micro-batches whose results are needed immediately by the next pipeline stage. For some of the micro-batches in the warm-up phase of 1F1B, the communication can be overlapped, but the degree is hard to model. We assume that they are perfectly overlapped.}.

\paragraph{Combining computation and communication.} 
As depicted in Fig. \ref{fig:model-parts}, it is possible to overlap the communication with the forward/backward pass computation. We use an intermediate variable $T_{cc}$ to denote the combination of computation and communication,
which is calculated as follows.

\hanyu{\emph{3D parallelism.} In 3D parallelism, the gradient synchronization of DP can be overlapped with the backward pass, whereas the communication for TP/PP cannot as it is on the critical path. We use a function $f_{overlap}^{k}(T_x, T_y)$ parameterized by $k$ to model the overlapping of two stages, where the fittable parameter $k$ represents the degree of the overlapping. Here we use $k_{sync}$ for the overlapping of DP and backward pass, thus we have $T_{cc}=T_{fwd}+f_{overlap}^{k_{sync}}(T_{bwd}, T_{comm\_dp})+T_{comm\_tp}+T_{comm\_pp}$, where the three communication times are calculated using the rule described above. To avoid distraction, we defer the detail of $f_{overlap}^{k}$ to Sec.~\ref{sec:model-together}.}

\emph{Gradient Accumulation.} When GA is used in DP, per-GPU gradients are aggregated locally over $a-1$ forward-backward passes before being synchronized across all GPUs during the $a^{th}$ pass. Therefore, the total backward propagation spans $a-1$ accumulation steps followed by the last step overlapped with the synchronization, that is, $T_{cc}=\revision{a\cdot} T_{fwd}+(a-1)\cdot T_{bwd}+f_{overlap}^{k_{sync}}(T_{bwd}, T_{comm\_dp})$.

\subsection{Modeling Optimizer and Offloading}
\label{sec:model-optimizer}

\paragraph{Modeling $T_{opt}$.}
The optimizer time $T_{opt}$ 
depends on the parameter size on each GPU, instead of the total parameter size, as the parameters are updated in parallel. We discuss each strategy as below.

\hanyu{\emph{3D parallelism or ZeRO-DP.} 3D parallelism and ZeRO-DP partition model parameters by the TP/PP size and DP size, respectively, thus we have $T_{opt}=k_{opt}\cdot P/x$, where $x$ represents $t \cdot p$ for 3D parallelism, and $d$ for ZeRO-DP.}

\emph{ZeRO-Offload.} Beyond the partitioning, ZeRO-Offload updates the partition each GPU owns directly on the CPU. Thus, we add a new fittable parameter to represent the CPU computation efficiency. Since CPU resources are used in parallel to jointly compute a single weight update, increasing the number of CPUs $c$ can also improve $T_{opt}$ under ZeRO-Offload, that is, $T_{opt}=k_{opt\_off}\cdot P/(d\cdot c)$.

\paragraph{Modeling $T_{off}$.}
$T_{off}$ represents the time specifically required by ZeRO-Offload, which is taken by the communication between CPU and GPU. ZeRO-Offload offloads the partitioned gradients to the CPU memory after computation and moves the parameter partitions back to the GPU after the parameter update. The communication volume for \hanyu{each data parallel GPU to the CPU is $P/d$ without mixed precision, thus we have $T_{off\_raw}=P/(d\cdot B_{pcie})$.}


\hanyu{In ZeRO-Offload, the offloading is also overlapped with the gradient synchronization and the optimizer step. We use an intermediate variable $T_{oo}$ to denote the combination of these parts. When using ZeRO-Offload, we have $T_{oo}=f_{overlap}^{k_{off}}(T_{comm\_dp},T_{off})+f_{overlap}^{k_{swap}}(T_{opt},T_{off})$; otherwise, we simply have $T_{oo}=T_{opt}$.}


\subsection{Putting It All Together}
\label{sec:model-together}

\hanyu{Combining the discussion in previous sections, we model the end-to-end iteration time as:
\begin{equation}\label{equation:iteration-offload}
\begin{split}
T_{iter}&=T_{cc}+T_{oo}+k_{const}
\end{split}
\end{equation}
where we use another fittable parameter $k_{const}$ to denote other constant overhead.}



\paragraph{Modeling overlapping.}
We use the function $f_{overlap}^k(T_x, T_y)$
to represent the total time spent by $x$ and $y$, considering the overlap between them. Taking the overlapping of $T_{bwd}$ and $T_{comm}$ as an example, if there is no overlap in data parallelism, they are combined as $T_{bwd}+T_{comm}$. If there is a perfect overlap, it should be $max(T_{bwd},T_{comm})$. A realistic value is somewhere in between these two extremes. To capture the overlapping, we borrow the definition from prior work~\cite{qiao2021pollux} as $f_{overlap}^k(T_x, T_y)=\left(T_{x}^{k}+T_{y}^{k}\right)^{\frac{1}{k}}$.
This formula has the property that the total time equals $T_{x}+T_{y}$ when $k=1$, and it smoothly transitions towards $max(T_{x},T_{y})$ as $k\xrightarrow{}\infty$.


\paragraph{\hanyu{Continuous} model fitting.}
The fittable parameters (listed in Table~\ref{tab:model_sum}) are fitted using throughput values collected from several sampled test runs using different resource allocations and execution plans. To fit such a $7$-tuple, we require at least seven data points before scheduling corresponding jobs. 
Considering that three parameters involve ZeRO-Offload ($k_{opt\_off}$, $k_{off}$, $k_{swap}$), the test runs should include three using this strategy.
We minimize the root mean squared logarithmic error (RMSLE) between Eq.~(\ref{equation:iteration-offload}) and the collected data triples. 
\revision{The model can also be updated online using metrics collected in real training runs when the prediction error exceeds a threshold.}
\hanyu{By continuously updating the model, \ourSys{} could fix potential prediction errors and the impact of such errors on scheduling decisions.}

\section{The \ourSys{} Scheduler}
\label{sec:scheduler}

\ourSys{} leverages the performance model and the reconfigurability of training jobs to improve job and cluster efficiency.
We begin with introducing the design goals, and then present the scheduling policy to achieve these goals.


\subsection{Design Goals}
\label{sec:scheduler-goals}

\ourSys{} is designed for the typical scenario of \emph{shared clusters}, where the resources are shared among multiple tenants (\eg, teams, departments) each owning a certain share of the resources (\eg, a resource \emph{quota}).
Each job is associated with a requested amount of resources.
Similar to existing systems, \ourSys{} classifies jobs in shared clusters into two categories \cite{wu2023transparent,apollo14,hived2020,antman2020,jeon2019analysis}. \emph{First} is \emph{guaranteed jobs} that consume certain amounts of resource quotas. Existing systems provide such jobs with the SLA guarantee that they are allocated at least the requested amounts of resources throughout their execution as long as the quota is enough. \revision{\emph{Second} is \emph{best-effort jobs} that do not occupy the resource quotas, and use free cluster resources opportunistically. These jobs can be preempted by the system (to avoid SLA violations for \emph{guaranteed} jobs).}

\ourSys{} follows the high-level principle of \emph{ensuring SLA for guaranteed jobs while improving resource utilization with best-effort jobs}.
Taking a step further, \ourSys{} redefines the scheduling goal to incorporate the new scheduling dimension of execution planning.
The first goal of \ourSys{} is to provide \textbf{performance guarantees} by ensuring that guaranteed jobs achieve at least the performance they would have with the requested resources and the original execution plans. Instead of guaranteeing exactly the resource amount, this new definition of SLA creates the opportunity for \ourSys{} to deliver the same or better performance with even \emph{fewer} resources by using better execution plans; such saved resources can further benefit other jobs. Based on the performance guarantees, the second goal is to \textbf{maximize cluster throughput}, \ie, the aggregated performance of all jobs (for both guaranteed and best-effort jobs).
We design a heuristic policy that continuously tunes the resource allocation and the execution plan for every job, based on the performance predictions by the performance model, to achieve such a global optimization.

\subsection{Scheduling Policy}
\label{sec:scheduler-polocy}

Our scheduling policy allocates GPU, CPU, and memory resources to jobs.
Multi-resource scheduling is analogous to the multi-dimensional bin packing problem, known as NP-hard~\cite{woeginger1997there}. Our problem is even more complex with a new dimension of execution planning.
Therefore, we aim to design a heuristic policy to achieve our goals efficiently.

\begin{figure}[t]
  \centering
\includegraphics[width=0.4\textwidth]{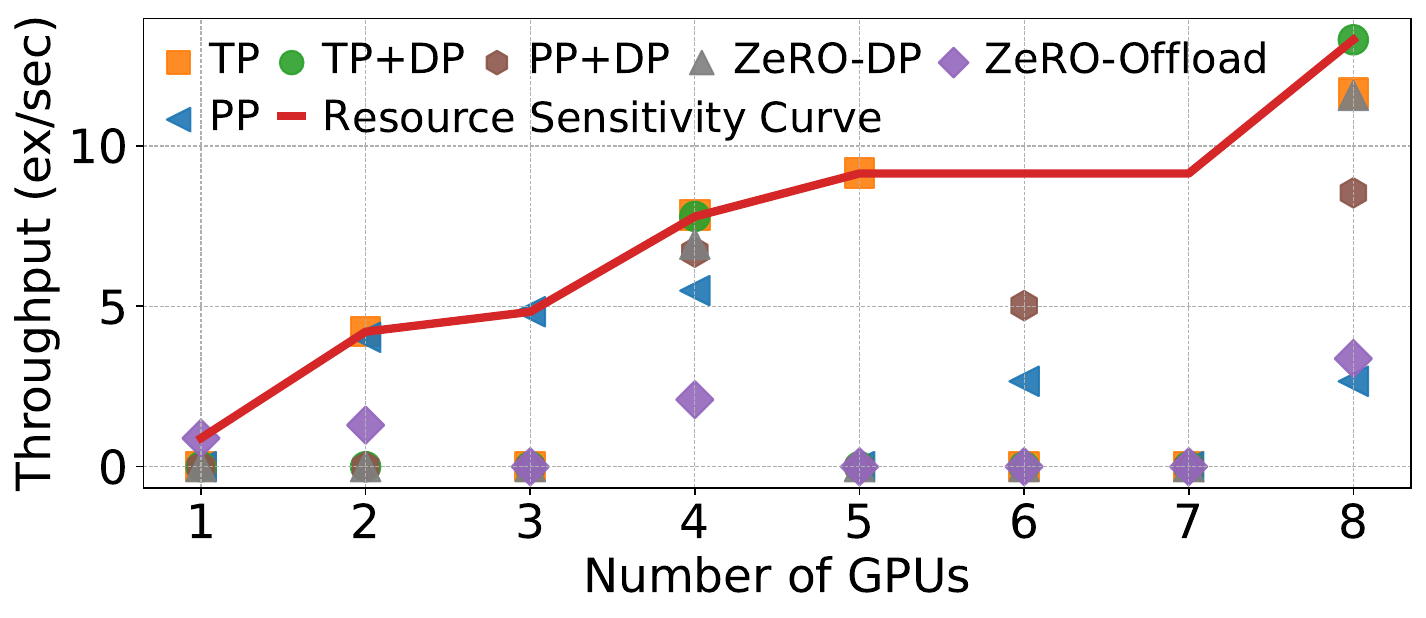}
  \caption{Resource (GPU) sensitivity curve of the \xinyi{GPT-2} model. Each point represents the throughput using the given GPU(s) with a certain execution plan. Only a few GPU numbers are valid (\newrevision{\ie, the data points with \emph{non-zero} job throughput}), due to the partitioning constraints of DP/TP/PP.}\label{fig:rscCurveStr}
\end{figure}

\paragraph{Resource sensitivity curves.}
\ourSys{} follows the principle of preferring jobs that benefit the most from the resources during scheduling to maximize cluster throughput. \ourSys{} understands and compares the benefits by building \emph{resource sensitivity curves} \revision{based on the performance model}. Intuitively, such a curve depicts the performance variation of a job when \revision{scaling} a certain type of resource (with other types fixed). Resource sensitivity curves also take execution planning into account, by only choosing the \emph{best plan} and the performance for each resource amount. \hanyu{\ourSys{} searches for the best execution plan for a job by enumerating the feasible plans and the performance predictions.} \revision{As shown in Fig.~\ref{fig:rscCurveStr}, the curve only connects the highest points along the x-axis that represent the best plans. The curve remains flat for invalid GPU numbers as it only considers the maximum throughput achievable within the given GPU range.}

Resource sensitivity curves help \ourSys{}'s scheduling policy in two ways. \emph{First,} the curves enable \ourSys{} to quickly pick the most sensitive jobs for allocation to maximize total throughput.
\emph{Second,} the curves make it possible for the scheduling algorithm to focus on the resource allocation with reduced complexity, while relying on the curves to provide corresponding best execution plans and performance predictions. Such a separation is beneficial because the curves can be computed in parallel or even prior to the scheduling, and then cached for reuse, improving the efficiency of the policy.

\paragraph{Scheduling algorithm design.}
To enforce performance guarantee, \ourSys{} in Algorithm~\ref{algorithm:policy} first searches for a minimum 
resource demand for each guaranteed job (denoted as $minRes$). The minimum  demand is not necessarily the user-requested resources;
instead, it is the fewest resources (possibly with a better execution plan) to achieve the performance of the original resource and plan.
The minimum demand also ensures that at least one plan can be trained without failures like GPU out-of-memory.
We require that the minimum demand \revision{should not exceed} the original in each dimension; if such a demand is not found, the original \revision{resource and plan} will be used.
For best-effort jobs, the minimum is $\Vec{0}$.

Our policy (function \texttt{Schedule}) is triggered whenever jobs are submitted or completed.
It first schedules the queuing jobs privileged to get scheduled immediately, \ie, the guaranteed jobs whose resource demands are within the tenant's remaining quota (lines \ref{line:previleged_start}-\ref{line:previleged_end}). We consider the quota usage of each job as its minimum demand to ensure a feasible allocation. The policy then continues to allocate 
resources, if any, to either schedule more best-effort jobs or to increase the allocation of running jobs (lines \ref{line:be_start}-\ref{line:be_end}).
\ourSys{} iterates over the nodes in the cluster to find a placement for each job (\texttt{ScheduleJob}). On each node, \ourSys{} searches for GPU and CPU to find an allocation to satisfy (possibly part of) the job's demand.
If at least the minimum demand is satisfied after the search,
we then select the execution plan \hanyu{given the found placement (\texttt{GetBestPlan}). Finally,
we allocate memory per the memory estimate of the assigned plan provided by the function \texttt{AllocMem} in the training framework (lines \ref{line:final_start}-\ref{line:final_end}).}
Note that we do not need to allocate memory during the search as it does not affect the performance.

{
\IncMargin{0.1em}
\begin{algorithm}[bt]
\SetKwData{Left}{left}\SetKwData{This}{this}\SetKwData{Up}{up} 
\footnotesize
\DontPrintSemicolon
\SetKwFunction{GetLowestSlopeOverMinJob}{GetLowestSlopeOverMinJob} \SetKwInOut{Input}
{Union}\SetKwFunction{Schedule}{Schedule} \SetKwInOut{Input}
{Union}\SetKwFunction{ScheduleJob}{ScheduleJob} \SetKwInOut{Input}
{Union}\SetKwFunction{SortJobsBySlope}{SortBySlope} \SetKwInOut{Input}
{Union}\SetKwFunction{GetMaxThroughputNode}{GetMaxThroughputNode} \SetKwInOut{Input}
{Union}\SetKwFunction{GetBestPlan}{GetBestPlan} \SetKwInOut{Input}
{Union}\SetKwFunction{AlignResRatios}{AlignResRatios} \SetKwInOut{Input}
{Union}\SetKwFunction{AllocMem}{AllocMem} \SetKwInOut{Input}
{input}\SetKwInOut{Output}{output}
\SetKw{Break}{break}
\SetKwProg{Fn}{Function}{:}{}
\Fn{\Schedule{$jobs$, $cluster$}}{
\For{$j \in jobs.privileged$}{\label{line:previleged_start}
$j.res, j.placement, j.plan = \ScheduleJob{j, cluster}$\;\label{line:previleged_end}
}
\For{$j \in$ \SortJobsBySlope{jobs.bestEffort $\cup$ jobs.running}}{\label{line:be_start}
$j.res, j.placement, j.plan = \ScheduleJob{j, cluster}$\;\label{line:be_end}
}
}
\Fn{\ScheduleJob{$j$, $cluster$}}{
\For{$n \in cluster.nodes$}{
$j.res$ $+=$ $n.freeRes$, $nodeRes=n.freeRes$\;\label{line:search_start}
\For{$resType \in \{GPU,CPU\}$}{
$\revision{\hat{j}}=\GetLowestSlopeOverMinJob{n, resType}$\;
\If(){$\revision{\hat{j}} == null$}{
\Break\;
}
\If{$j$.slope($resType$) $>$ \revision{$\hat{j}$}.slope($resType$) || $j.res[resType] < j.minRes[resType]$}{\label{line:permit_shrink}
$\revision{\hat{j}}.res$ $-=$ $\Delta{r}$, $j.res$ $+=$ $\Delta{r}$, $nodeRes$ $+=$ $\Delta{r}$\;
}
\Else{
\Break\;\label{line:search_end}
}
}
\If{$nodeRes > \Vec{0}$}{ \label{line:record_start}
$j.placement$.append($\{n, nodeRes\}$)\;
}\label{line:record_end}
}
\If{$j.res>=j.minRes$}{\label{line:final_start}
$plan=$\GetBestPlan{$j,j.placement$}\;
$success=$\AllocMem{$j.res,plan$}\;
\If{$success$}{
\KwRet $j.res$, $j.placement$, $plan$\;\label{line:final_end}
}
}
\KwRet $null$, $null$, $null$\;
}
\caption{\ourSys{} Scheduling Policy}
\label{algorithm:policy} 
\end{algorithm}
\DecMargin{0.1em} 
}

\ourSys{} evaluates the gains of allocating resources to different jobs according to the \emph{slopes} of their resource sensitivity curves. \revision{We define the slope, which is specific to each resource type of different jobs, as the throughput change per unit variation in the number of (pre-)allocated resources.}
On each node, besides the free resources, \ourSys{} is also allowed to ``shrink'' other jobs to reclaim and reallocate resources (lines \ref{line:search_start}-\ref{line:search_end}). \newrevision{After that, \ourSys{} records the resource allocation results for the current scheduled job on each node (lines \ref{line:record_start}-\ref{line:record_end}).}
Specifically, \ourSys{} always shrinks the least sensitive job, \ie, the one with the \emph{lowest} slope (\texttt{GetLowestSlopeOverMinJob}, where ``OverMin'' means that the job must be over its own minimum demand). Such a reallocation is permitted in \revision{two} cases (line \ref{line:permit_shrink}): (1) the job to shrink has a slope lower than the job to schedule, thus the reallocation will increase total throughput; or (2) the job to schedule has yet to reach its minimum demand, then a reallocation that decreases total throughput is also acceptable to meet the performance guarantee.
\ourSys{} reallocates a unit of the resource ($\Delta{r}$) repeatedly until further reallocation is not allowed.
Shrinking a job to $\Vec{0}$ results in a preemption (possible only for best-effort jobs), which will return to the queue.


Similarly, when choosing best-effort or running jobs for allocation, \ourSys{} also prefers those with the \emph{highest} resource sensitivity curve slopes for the most throughput improvement (\texttt{SortBySlope} at line \ref{line:be_start}). Considering multiple resource dimensions, here we do a greedy sort that compares the slopes of GPUs and then CPUs.
\revision{Unscheduled} guaranteed jobs do not need such a sort as they are chosen with respect to the quotas.
\newrevision{Although the scheduling policy of \ourSys{} encourages those jobs that can utilize resources efficiently, to prevent job starvation and inaccurate estimation in the performance model, best-effort jobs get scheduled occasionally when the queuing delay exceeds a threshold~\cite{gu2019tiresias}.}


\revision{In particular, \ourSys{} supports distributed training by placing a job on multiple nodes during the search. As our performance model explicitly considers the inter-node bandwidth ($B_{inter}$ in Table~\ref{tab:model_sum}), the resource sensitivity curves can capture the performance variation when jobs become distributed.}

\begin{table*}[ht]
\centering
\caption{\hanyu{Performance \faye{prediction} errors of \faye{representative Transformer-based} models used in our evaluation. TP+PP: adjusting TP/PP sizes with DP$=1$; DP+TP+PP: adjusting DP with fixed TP/PP sizes. ``/'' denotes the infeasible plan due to OOM.}}
\footnotesize{
\begin{tabular}{cccc|cccccccc}
\toprule

\multirow{2}{*}{\textbf{Model}}  & \multirow{2}{*}{\textbf{Size}}  & \multirow{2}{*}{\textbf{Dataset}} &  \multirow{2}{*}{\textbf{\# GPUs predicted}} & avg.      & max.     & avg.     & max.     & avg.           & max.          & avg.          & max.     \\

\cmidrule(r){5-12}

 & & & & \multicolumn{2}{c}{DP} & \multicolumn{2}{c}{GC} & \multicolumn{2}{c}{ZeRO-DP+GA} & \multicolumn{2}{c}{ZeRO-Offload}  \\
\cmidrule(r){1-5} \cmidrule(r){5-6} \cmidrule(lr){7-8} \cmidrule(lr){9-10} \cmidrule(lr){11-12}
ViT~\cite{dosovitskiy2021image}   & $86$M     &  ImageNet-1K~\cite{imagenet09} & $[1-8]$   & $3.63\%$    & $6.83\%$   & $2.59\%$   & $6.19\%$   & $4.23\%$         & $6.67\%$        & $3.00\%$           & $8.32\%$        \\
RoBERTa~\cite{liu2019roberta} & $355$M   & WikiText-2~\cite{merity2016pointer} & $[1-8]$ & $2.21\%$    & $4.37\%$   & $3.36\%$   & $4.29\%$   & $3.59\%$         & $6.71\%$        & $7.42\%$        & $10.44\%$       \\
BERT~\cite{bert19}  & $336$M    &   Wikipedia~\cite{wikidump} & $[1-8]$  & $5.27\%$    & $8.32\%$   & $4.90\%$    & $7.27\%$   & $3.7\%$     
     & $6.90\%$         & $6.37\%$        & $8.62\%$        \\
 \addlinespace
\cline{5-12}
\addlinespace

 &  & & & \multicolumn{2}{c}{TP+PP} & \multicolumn{2}{c}{DP+TP+PP} & \multicolumn{2}{c}{ZeRO-DP+GA} & \multicolumn{2}{c}{ZeRO-Offload+GC} \\
\cmidrule(r){5-6} \cmidrule(lr){7-8} \cmidrule(lr){9-10} \cmidrule(l){11-12}
T5~\cite{raffel2020exploringt5}     & $1.2$B     &  Wikipedia~\cite{wikidump} &  $[1-32]$  & $3.18\%$    & $8.24\%$   & $2.41\%$   & $9.55\%$   & $6.71\%$         & $9.55\%$        & $4.37\%$        & $6.34\%$        \\
GPT-2~\cite{Radford2019LanguageMAGPT2} & $1.5$B     & Wikipedia~\cite{wikidump} & $[1-30]$ & $2.39\%$    & $3.08\%$   & $2.80\%$    & $4.15\%$   & $2.52\%$         & $3.86\%$        & $5.52\%$        & $8.90\%$         \\
LLaMA-2-$7$B~\cite{touvron2023llama2}  & $7$B      & WuDaoCorpora~\cite{yuan2021wudaocorpora} & $[1-64]$ & $1.90\%$     & $2.90\%$    & $4.70\%$    & $9.45\%$   & /              & /             & $4.09\%$        & $6.38\%$        \\
LLaMA-$30$B~\cite{touvron2023llama} & $30$B    & WuDaoCorpora~\cite{yuan2021wudaocorpora} &  $[12-64]$  & $4.29\%$    & $8.52\%$   & $6.15\%$   & $9.69\%$   & /              & /             & /             & /             \\
\bottomrule
\end{tabular}
}
\label{tab:models}
\end{table*}

\paragraph{Incorporating reconfiguration penalty.} \ourSys{} adopts the common checkpoint-resume method when switching execution plans~\cite{gandiva2018}, where the overhead can be limited through reconfiguration frequency. 
\ourSys{} considers to reconfigure a job only if the value of ${(T-N\cdot \delta)}/{T}$ exceeds a threshold. $T$ is the aggregated training time of the job, and $N$ is the reconfiguration times. $\delta$ is the checkpoint-resume cost of the job, and the threshold is set as $0.97$ empirically.

\section{Implementation}
\label{sec:implementation}

We implement a prototype of the \ourSys{} scheduler on Kubernetes~\cite{k8s} in Python. The scheduler uses Kubernetes APIs to watch the creation and completion of pods (\ie, Docker containers of training workers) and cluster resource status.
The lifecycle of training jobs and pods is managed by Kubeflow \cite{kubeflow}.
In each scheduling round, the scheduler runs its scheduling policy and applies the resultant allocations by (re-)launching the training jobs.

We use two popular PyTorch-based large-model training frameworks, DeepSpeed~\cite{jeff2020deepspeed} and Megatron~\cite{megatron2019} for the model training (PyTorch 1.12, DeepSpeed 0.9.2, and Megatron-DeepSpeed v2.4). With PyTorch providing an official launching API to start distributed training, \ourSys{} can (re-)configure training jobs with different execution plans by simply making minor modifications to the launching command (without modifying the model or the framework codes). When re-launching a job, \ourSys{} instructs the job to save a checkpoint before exiting, and then the job resumes from the checkpoint after the restart. 
\revision{\ourSys{} relies on the inherent capability of DeepSpeed and Megatron to estimate the memory consumption and correct modeling for inaccurate predictions. }
\revision{For CPU resources, each training process is bound to the allocated CPU cores to ensure the dedicated usage, which benefits the training performance under ZeRO-Offload.}
\revision{As for profiling, \ourSys{} measures the bandwidths of different link types, \eg, NVLink and PCIe.}

\section{Evaluation}
\label{sec:evaluation}

We evaluate \ourSys{} using experiments on a 64-GPU cluster and trace-driven simulations. 
\hanyu{
The cluster is comprised of $8$ servers, each equipped with $8$ NVIDIA A800 GPUs ($80$ GB), $96$ vCPUs, $1,600$ GB memory, $400$ GB/s NVLink bandwidth, and $100$ GB/s RDMA network bandwidth.
We use seven representative Transformer-based models of various scales as listed in Table~\ref{tab:models}.
}
Overall, our key findings include:
\begin{packed_itemize}
     \item \ourSys{} significantly improves job and cluster efficiency in the 64-GPU cluster, achieving up to $3.2\times$ JCT gain over state-of-the-art reconfigurability-agnostic systems.
     \item \ourSys{} enforces the performance guarantees via job reconfiguration, achieving $1.7\times$ JCT gain for guaranteed jobs
     compared to using exact resource guarantees. 
     \item \ourSys{} shows increasing JCT gains (from $2.6\times$ to $3.4\times$) with larger proportions of large models, which shows the potential of \ourSys{} in the large-model era.
\end{packed_itemize}

\subsection{Performance Model Validation}
\label{sec:model-validation}

\hanyu{
We validate our performance model on the seven deep learning models in Table~\ref{tab:models} using up to $64$ A800 GPUs. For each model, we fit the performance model using the minimum set of $7$ profiled data points.
We then predict the performance for $20$ unseen configurations, \ie, $4$ execution plans each with $5$ multi-resource allocations or placements. For models having fewer than $1$B parameters, we predict DP, GC, ZeRO-DP+GA, and ZeRO-Offload using $1$ to $8$ GPUs. For larger models, we also predict 3D parallelism with changing DP/TP/PP sizes using more GPUs. Table~\ref{tab:models} shows the average and maximum errors of the predictions for each execution plan of each model. The average and maximum errors are up to $7.4\%$ and $10.4\%$, respectively, showing good prediction quality. Despite the errors, \ourSys{} also continuously fits the model after a job is launched, further mitigating the errors.
}

\subsection{Micro-benchmarks}
\label{sec:evaluation-micro}

This section constructs several cases to demonstrate the reconfiguration ability of \ourSys{} and its benefits. 



\begin{figure}[tb]
  \centering
  \includegraphics[width=0.35\textwidth]{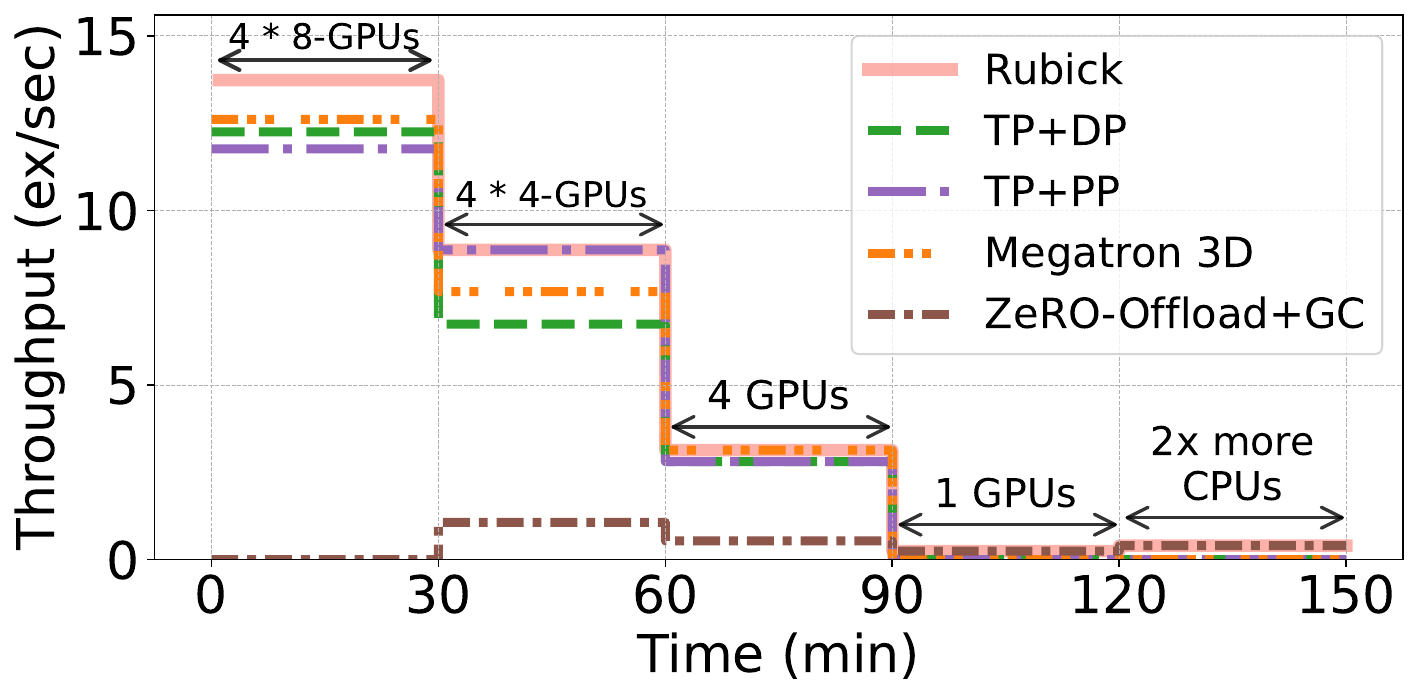}
  \caption{\xinyi{Reconfiguration for a LLaMA-2-7B job by \ourSys{}. See the caption of Fig.~\ref{fig:moti_throughput} for the definitions of the plans.}}\label{fig:rsc_limit}
\end{figure}

\paragraph{Adapting to changing resource limits.}
In this experiment, we train a LLaMA-2-7B model while continuously decreasing the limits of available resources.
As shown in Fig.~\ref{fig:rsc_limit}, although the best plans vary over time, \ourSys{} \emph{always chooses the best}.
Firstly, the model is trained across $4$ servers each with $8$ A800 GPUs.
\hanyu{
\ourSys{} chooses an optimal 3D-parallel configuration (DP=$4$, PP=$2$, TP=$4$), which is even better than those found by other simple 3D parallelism tuning strategies shown by the other lines in Fig.~\ref{fig:rsc_limit}.
We then decrease the GPUs to $16$ ($4*4$) and $4$, and \ourSys{} still uses the best 3D-parallel configurations.
}
When the number of GPUs is reduced to $1$, the GPU memory estimator in \ourSys{} instructs to choose ZeRO-Offload, the only feasible plan with only one GPU available. 
Upon shifting to ZeRO-Offload, \ourSys{} also increases the memory allocation to satisfy its demand.
Finally, we double the available CPU resources,
and \ourSys{} acquires $1.7\times$ speedup by allocating the CPUs to accelerate the parameter updates.



\paragraph{Maximizing throughput across jobs.}
To highlight \ourSys{}'s ability to maximize throughput considering jobs' resource sensitivity,
we compare it with a simple scheduler that equalizes resource allocation across jobs. We also allow the simple scheduler to reconfigure execution plans, thus focusing on the difference between scheduling policies.

\begin{figure}[tb]
  \centering
  \includegraphics[width=0.42\textwidth]{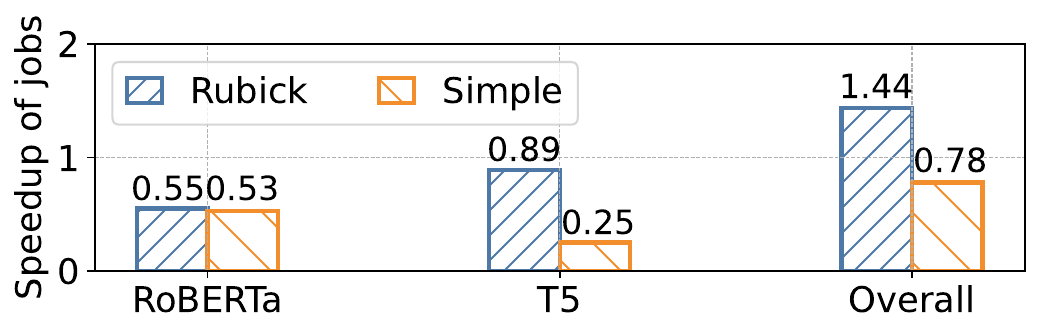}
  \caption{\xinyi{Throughput \revision{improvement} across two jobs.}}
  \label{fig:micro_colocate_tpt}
\end{figure}

We submit a RoBERTa job and a T5 job to a cluster of \xinyi{ $4$ A800 GPUs. }
\revision{To quantify the total throughput of the jobs, we normalize the throughput of each job as a factor of speedup improvement to a rigid execution plan on static resources~\cite{qiao2021pollux}. The baseline refers to the performance on \xinyi{$4$} GPUs.} 
\revision{As shown in Fig.~\ref{fig:micro_colocate_tpt},} the simple scheduler allocates \xinyi{$2$} GPUs to each job evenly, and reconfigures T5 and RoBERTa to use \xinyi{ZeRO-Offload and ZeRO-DP}, respectively, which results in a total \revision{speedup} of \xinyi{$0.78$}.
In comparison, \ourSys{} identifies that the T5 model can \newrevision{achieve} more performance gain from more GPUs than RoBERTa. \ourSys{} therefore allocates \xinyi{$3$} GPUs to T5 and \xinyi{$1$} GPU to RoBERTa, and reconfigures them to use \xinyi{TP with GA and DP with GA}, respectively.
This allocation leads to a total \revision{speedup} of \xinyi{$1.44$}, with \xinyi{$85\%$} \newrevision{performance} improvement as compared to the simple allocation.

\begin{figure}[t]
  \centering
  \includegraphics[width=0.47\textwidth]{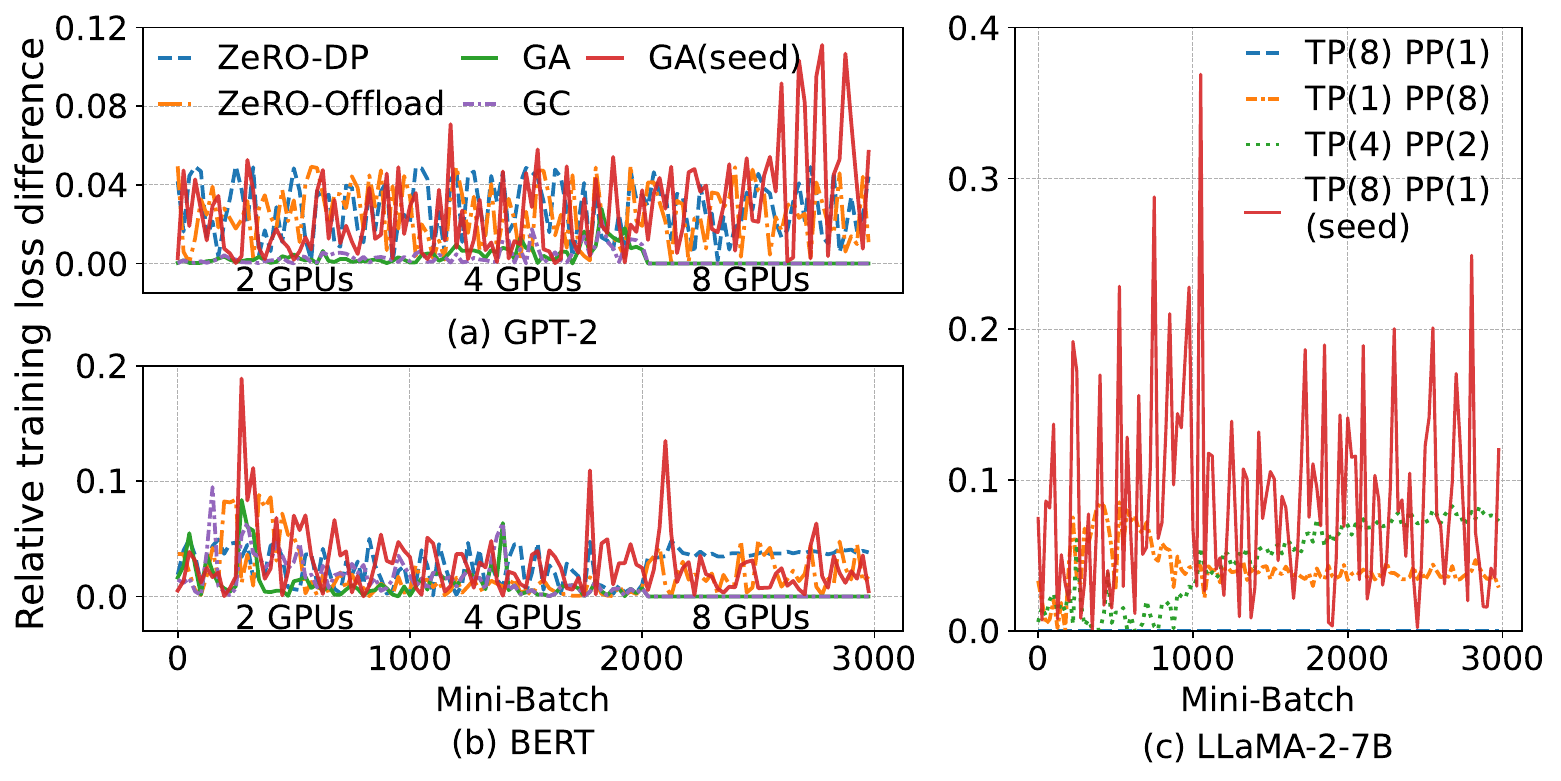}
  \caption{\revision{\xinyi{Relative loss difference during reconfiguration.}}}\label{micro:loss}
\end{figure}

\paragraph{\revision{Accuracy during reconfiguration.}}
\hanyu{\ourSys{} keeps the global batch size unchanged during resource scaling and reconfiguration, thus not affecting training accuracy by design. To validate this, we compare the training losses of different resource allocations and execution plans to that without reconfiguration but using a different random seed, which represents an acceptable range of accuracy variance due to randomness. We train GPT-2 and BERT using $2$/$4$/$8$ GPUs and LLaMA-2-7B using $8$ GPUs with different execution plans.
Each experiment trains the model for $3,000$ mini-batches. We choose one of the resource-plan combinations as the accuracy baseline and plot the relative difference curves of the others (\ie, GA on $8$ GPUs for GPT-2 and BERT, TP$=8$ and PP$=1$ for LLaMA-2-7B). Curves denoted with ``seed'' use a different random seed for a certain execution plan. As shown in Fig.~\ref{micro:loss}, the train losses of different resources/plans fluctuate mostly within the range of changing random seeds. We also compare the validation and test losses after $3,000$ mini-batches. As shown in Table~\ref{tab:micro-loss}, the maximum loss differences of reconfiguration on train, validation, and test datasets are always smaller than those of altering seeds, showing the negligible impact on training accuracy of \ourSys{}.}

\subsection{Cluster Experiments}
\label{sec:evaluation-cluster}

\paragraph{Methodology.} 
We compare \ourSys{} with three state-of-the-art schedulers: (1) Sia~\cite{subramanya2023sia}, which tunes GPU numbers \hanyu{by adjusting the DP size\footnote{Despite the claim in their paper of supporting 3D parallelism, Sia's open-source artifact~\cite{siaartifact} only supports pure DP jobs. Their evaluation tested 3D-parallel jobs only with a small-scale simulation. Adding 3D-parallelism support in Sia's artifact is non-trivial; we implemented the claimed scaling approach of Sia, \ie, scaling DP for 3D-parallel jobs, in another baseline \ourSys{}-R. In our experiments for Sia, if a model cannot run using DP (even when ZeRO/GA/GC), the job fallbacks to a feasible 3D-parallel plan with the resource scaling disabled.}
together with their hyper-parameters (which influence their accuracy) to improve the ``goodput'', \ie, to reduce the ``time-to-accuracy''.}
(2) Synergy~\cite{synergy22}, which tunes CPU-memory allocation for \revision{GPU jobs with fixed GPU numbers}.
(3) AntMan~\cite{antman2020}, a multi-tenant scheduler that provides the concepts of guaranteed and best-effort jobs similar to \ourSys{}.
We also establish three variants of \ourSys{} for a break-down comparison: 
(1) \emph{\ourSys{}-E} only reconfigures execution plans with fixed resources. (2) \emph{\ourSys{}-R} only reallocates resources with fixed execution plans. \hanyu{For 3D-parallel jobs, \ourSys{}-R uses the same approach of Sia that changes the DP size when scaling GPUs.} (3) \emph{\ourSys{}-N} does neither of them, and only applies \ourSys{}'s scheduling policy.

\begin{table}[t]
\centering
\caption{Maximum loss differences of reconfiguration (``Rcfg.'') and changing random seeds (``Seed'').}\label{tab:micro-loss}
\footnotesize{
\begin{tabular}{c|c|c|c|c|c|c}
\toprule[1pt]
\multicolumn{1}{c|}{\multirow{2}{*}{Model}} & \multicolumn{2}{c|}{Train} & \multicolumn{2}{c|}{Validation} & \multicolumn{2}{c}{Test} \\ \cline{2-7}
\multicolumn{1}{c|}{}                  & Rcfg.        & Seed       & Rcfg.          & Seed          & Rcfg.       & Seed       \\ \midrule
GPT-2                                 & $0.05$         & $0.11$       & $0.08$           & $0.09$           & $0.10$         & $0.21$      \\
BERT                                  & $0.10$          & $0.19$       & $0.10$            & $0.10$           & $0.38$         & $0.40$        \\
LLaMA-2-7B                            & $0.08$         & $0.37$       & $0.07$           & $0.41$          & $0.10$         & $0.11$      \\
\bottomrule[1pt]
\end{tabular}
}
\end{table}

We construct synthetic traces by down-sampling the busiest $12$ hours in a GPU cluster trace published by Microsoft~\cite{jeon2019analysis}, proportionally to the cluster sizes. The sampled trace contains $406$ jobs, each with a submission time, number of GPUs, and duration. 
\hanyu{
For each job, we select a model from Table~\ref{tab:models} randomly. In case the original GPU number is infeasible for the model, we use a feasible one and change the duration accordingly to keep the same GPU hours of the job.}
For all schedulers except Sia, we translate the job duration to a target number of mini-batches using the measured throughput of model with the GPU number. For Sia, to meet its goal of reducing time-to-accuracy, we assign a target evaluation accuracy (instead of fixed mini-batches) to each job, measured by running the model for the specified duration. 

We build three variants of the sampled trace for different specific scenarios. (1) \emph{Base trace},
\hanyu{which randomly assigns an initial execution plan to each job from all feasible plans given the GPU number. For ViT, RoBERTa, BERT, and T5, we disable TP and PP as they are mostly unnecessary for these relatively small models. For the other models, we include all the feasible 3D-parallel configurations in the candidate plans.
}
(2) \emph{Multi-tenant trace (MT)}, a multi-tenant version of the base trace.
This trace sets up two tenants, Tenant-A with a quota of $64$ GPUs, and Tenant-B with no quota, and randomly dispatches jobs to them. Jobs from Tenant-A/B are all guaranteed/best-effort, respectively.
\hanyu{(3) \emph{Best-plan trace (BP)}, which replaces the random execution plans in the base trace with the best plans of the corresponding jobs given the initial resource amounts.}



\paragraph{End-to-end comparison.}
As shown in Table~\ref{tab:cluster-experiment}, \ourSys{} consistently achieves the shortest average and P99 job completion times (JCT) and makespan using different traces.
\hanyu{
With the base trace, \ourSys{} achieves up to $3.2\times$, $1.9\times$, and $1.4\times$ improvement compared to Sia and Synergy on average JCT, P99 JCT, and makespan, respectively.
Sia, despite GPU scaling along the DP dimension, has limited support for advanced training strategies beyond DP. It cannot scale 3D-parallel jobs with TP/PP; also, its performance model cannot capture behaviors of ZeRO/GC, and ignores multi-resource allocations beyond GPUs.
\ourSys{} outperforms Sia by $2.6\times$ in average JCT, highlighting the advantage of \ourSys{}'s full reconfigurability on a wide range of execution plans and multiple resources.
\ourSys{} also outperforms Synergy by $3.2\times$ in average JCT because Synergy does not consider execution planning during its multi-resource allocation.
With the best-plan (BP) trace, Sia and Synergy perform better with better execution plans. Even in such a case, \ourSys{} still shows $1.9\times$ and $2.4\times$ average JCT gains over Sia and Synergy, respectively, because the assigned plan is the best only for the initial resource allocation; \ourSys{} can further reconfigure the plan together with the resource scaling, showing the necessity of adapting the execution plans to the resource variations.
}


\begin{table}[t]
\renewcommand{\arraystretch}{1.2} 
\centering
\caption{\xinyi{$64$-GPU cluster experiments. ``All'', ``Guar.'', and ``BE'' stand for all, guaranteed, and best-effort jobs, respectively.}}
\footnotesize
\begin{tabular}{c|c|c|c|c|c}
\toprule[1pt]
\multicolumn{1}{c|}{\multirow{2}{*}{\textbf{Trace}}}    & \multicolumn{2}{c|}{\multirow{2}{*}{\textbf{Scheduler}}} & \multicolumn{2}{c|}{\textbf{JCT (h)}} & \multirow{2}{*}{\textbf{\makecell[c]{Makespan\\(h)}}} \\
\cline{4-5}
&\multicolumn{2}{c|}{}&Avg.&P99& \\
\midrule[1pt]
\multirow{6}{*}{Base}          & \multicolumn{2}{c|}{\ourSys{}}    & $\mathbf{0.96}$ ($1\times$)   & $\mathbf{7.1}$($1\times$)   & $\mathbf{15.3}$ ($1\times$)                \\ \cline{2-6} 
                               & \multicolumn{2}{c|}{Sia}    & $2.5$ ($2.6\times$)                    & $12.2$ ($1.7\times$)  & $18.8$      ($1.23\times$)          \\ \cline{2-6} 
                               & \multicolumn{2}{c|}{Synergy}   & $3.1$($\mathbf{3.23}\times$)                    & $13.5$ ($\mathbf{1.9}\times$)  &  $21.5$  ($1.4\times$)           \\ \cline{2-6} 
                               & \multicolumn{2}{c|}{\ourSys{}-E}    & $2.4$ ($2.5\times$)                    & $10.9$ ($1.5\times$)    & $20.2$  ($1.32\times$)                \\ \cline{2-6} 
                               & \multicolumn{2}{c|}{\ourSys{}-R}    & $1.6$ ($1.67\times$)         & $9.9$ ($1.39\times$)     & $19.8$ ($1.29\times$)                    \\ \cline{2-6} 
                               &  \multicolumn{2}{c|}{\ourSys{}-N}    & $3.1$ ($\mathbf{3.23}\times$)    & $12.8$ ($1.8\times$)     & $22$ ($\mathbf{1.44}\times$)                    \\ \midrule[1pt]
\multirow{3}{*}{BP} & \multicolumn{2}{c|}{\ourSys{}}    & $\mathbf{0.96}$ ($1\times$)   & $\mathbf{7.1}$($1\times$)   & $\mathbf{15.3}$ ($1\times$)                \\ \cline{2-6} 
&\multicolumn{2}{c|}{Sia}    & $1.8$ ($1.88\times$) & $9$  ($1.27\times$)   & $16.5$ ($1.08\times$)                   \\ \cline{2-6} 
                               & \multicolumn{2}{c|}{Synergy}   & $2.3$ ($\mathbf{2.37\times}$)        & $10.8$ ($\mathbf{1.5\times}$) & $20.5$ ($\mathbf{1.34}\times$)                  \\
\midrule[1pt]
\multirow{6}{*}{MT} & \multirow{3}{*}{\ourSys{}}   & All   & $\mathbf{1.1}$ ($1\times$)  &$\mathbf{11.4}$ ($1\times$)   & \multirow{3}{*}{$\mathbf{17.9}$ ($1\times$)}                   \\ \cline{3-5} 
& & Guar.   & ${0.85}$ ($1\times$)  &$10.9$ ($1\times$)   &  \\ \cline{3-5} 
&  & BE   & $1.34$ ($1\times$)  &$11.8$ ($1\times$)   &                    \\ \cline{2-6} 
                               & \multirow{3}{*}{AntMan}  & All   & $1.75$ ($\mathbf{1.6\times}$)    & $13.4$ ($\mathbf{1.2\times}$)       &\multirow{3}{*}{$19.6$ ($\mathbf{1.28\times}$)}                 \\ \cline{3-5} 
& & Guar.   & $1.41$ ($1.65\times$)  &$11.7$ ($1.1\times$)   &                  \\ \cline{3-5} 
&  & BE   & $2.1$ ($1.56\times$)  &$14.1$ ($1.2\times$)   &                    \\ 
                               
\bottomrule[1pt]
\end{tabular}
\label{tab:cluster-experiment}
\end{table}

We compare \ourSys{} with AntMan using the multi-tenant (MT) trace to evaluate the SLA guarantees.
Overall, \ourSys{} outperforms AntMan by $1.6\times$ and $1.3\times$ in average JCT and makespan, respectively.
Their key difference is that AntMan guarantees the requested resources, whereas \ourSys{} guarantees the corresponding performance during reconfiguration.
\hanyu{For guaranteed jobs, \ourSys{} improves the average JCT by $1.7\times$, showing that \ourSys{} not only guarantees, but also improves their efficiency with better execution plans. Similarly, \ourSys{} also shows $1.6\times$ JCT gain for best-effort jobs.}

\paragraph{Break-down study.}
We use the base trace to compare \ourSys{} to the three variants, \ourSys{}-E/R/N, to understand the sources of improvements. As shown in Table~\ref{tab:cluster-experiment}, by reconfiguring execution plans or reallocating resources solely, \ourSys{}-E and \ourSys{}-R improve the average JCT compared to \ourSys{}-N by $1.3\times$ and $1.9\times$, respectively.
This demonstrates that these two approaches are already powerful weapons even used separately; however, the complete \ourSys{} still shows $2.5\times$ and $1.7\times$ extra improvements, further highlighting the necessity of the combination of them.

\paragraph{\revision{System overheads.}}
\hanyu{For each job, the average time spent on reconfiguration is $78$ seconds, and the total reconfiguration time accounts for $1\%$ in total GPU hours across all experiments.}
For each model in Table~\ref{tab:cluster-experiment}, workload profiling only takes an average of $210$ seconds to collect performance values from $7$ sampled tests on an $8$-A800 server.

\begin{figure}[t]
  \centering
    \includegraphics[width=0.49\textwidth]{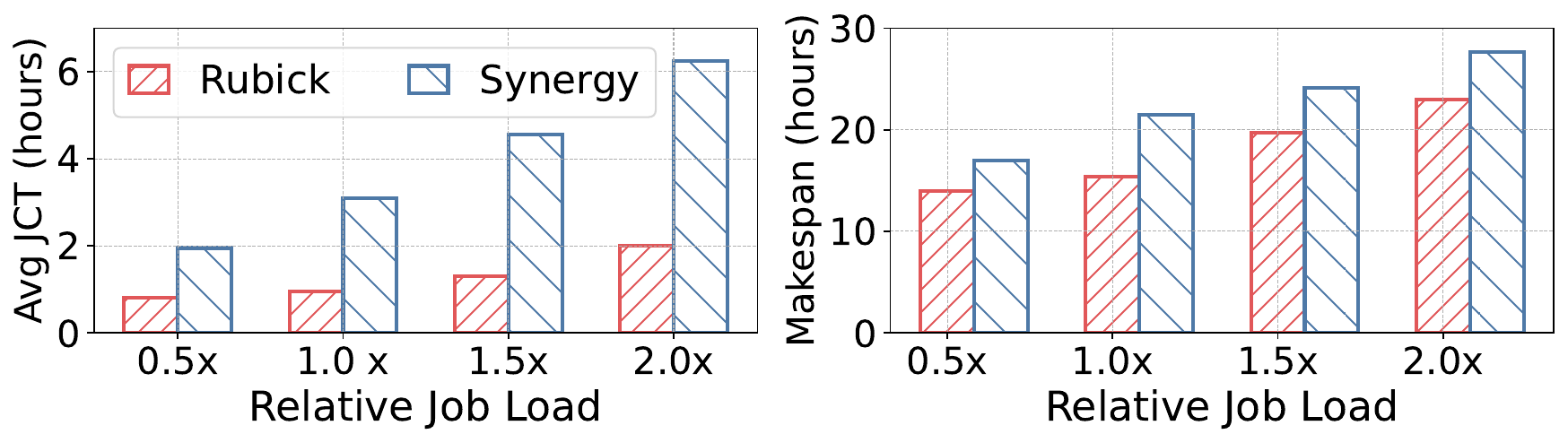}
  \caption{Performance vs. cluster load.}\label{fig:simu_load}
\end{figure}

\begin{figure}[t]
  \centering
    \includegraphics[width=0.49\textwidth]{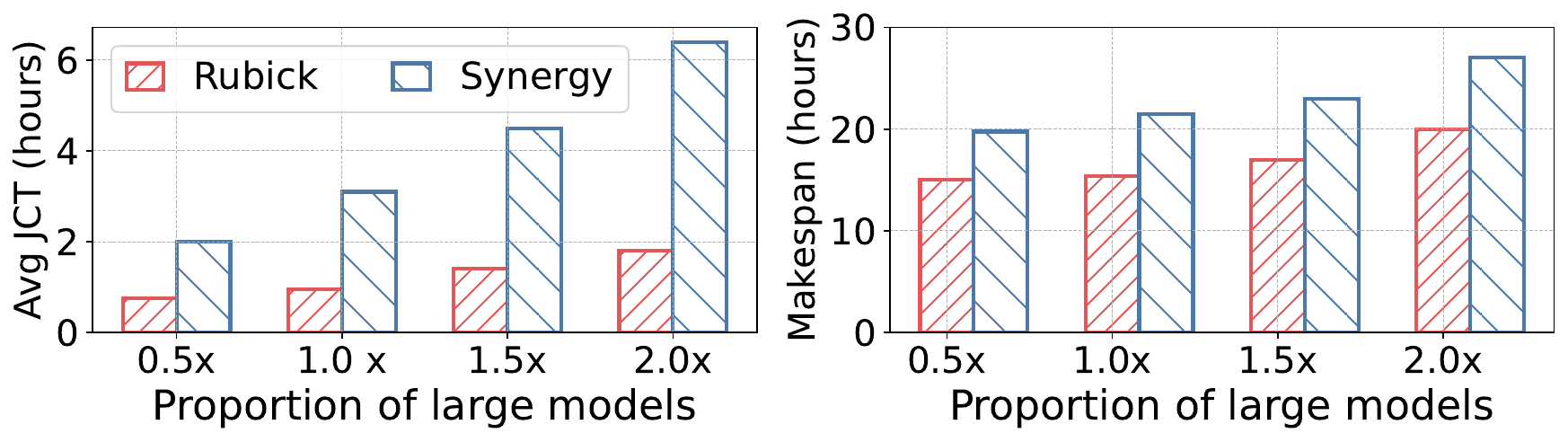}
  \caption{\xinyi{Performance vs. proportion of large models.}}\label{fig:simu_model}
\end{figure}

\subsection{Simulations}
\label{sec:evaluation-simulations}

We use simulations to evaluate \ourSys{} with various settings to reveal the factors that affect its behavior and performance.
We build a discrete-time cluster simulator, and use real performance measurement to estimate the execution time of the jobs. 
We replayed the cluster experiments in Sec.~\ref{sec:evaluation-cluster} with the simulator and the max error of average JCT was \xinyi{$6.9\%$}.



\paragraph{Performance with varying cluster load.}
We vary the load of the traces with different down-sampling rates.
Fig.~\ref{fig:simu_load} shows the performance of \ourSys{} and Synergy with increasing load ($1\times$ corresponds to the original sampling rate). \ourSys{} consistently outperforms Synergy under all loads, with up to \xinyi{$3.5 \times$ and $1.4\times$} improvements for JCT and makespan. In general, higher loads lead to more gains of \ourSys{} because the improvements are accumulated across all queuing jobs.


\paragraph{Performance with varying model size distribution.}
The job reconfigurability in \ourSys{} enables even larger ranges of resource availability feasible for training a model. This property is especially beneficial for large models because
they have the opportunity to start training earlier with fewer GPUs. 
We compare the performance of \ourSys{} and Synergy with an increasing proportion of large models (\xinyi{LLaMA-2-7B and LLaMA-30B}) in the trace.
Fig.~\ref{fig:simu_model} shows that \ourSys{}'s advantage keeps increasing with more large models, with the JCT gain ranging from \xinyi{$2.6\times$ to $3.4\times$}.
We view such increasing benefits with an increased number of large models 
as a nice property of \ourSys{} that would be appreciated, 
as this is exactly the developing trend today.

\section{Related Work}
\label{sec:related}

\paragraph{Parallelization strategies and optimizations.}
To facilitate the training of large models, tensor parallelism~\cite{lepikhin2021gshard} and pipeline parallelism~\cite{huang2019gpipe,narayanan2019pipedream} are proposed to partition the model across GPUs. To optimize the GPU memory usage, DeepSpeed~\cite{jeff2020deepspeed} and ZeRO series~\cite{rajbhandari2020zero, ren2021zero, rajbhandari2021zero} further  partition the model and offload its computation, weights, and optimizer states to the main memory. 
Gradient checkpointing~\cite{chen2016training, jain2020checkmate} trades the computation for GPU memory by recomputing tensors from checkpoints. 
These prior techniques above can offer numerous options for job execution plans to \ourSys{}.
To generate a unified job execution plan, Alpa~\cite{zheng2022alpa} automates inter-operator (\ie, model and pipeline) parallelism and intra-operator (\ie, data and tensor) parallelism.
Similarly, Unity~\cite{unger2022unity} adopts a generic and extensible approach to optimize the job execution plan with both parallel strategies and graph substitutions.
Whale~\cite{jia2022whale} automatically \revision{decides} parallel strategies by considering the capacities of heterogeneous GPUs. 
The aforementioned studies presume servers are committed to complete job execution, whereby an optimal job plan is initially searched for and subsequently executed in a static manner.
\hanyu{
Currently, \ourSys{} focuses on showing the potential of co-optimizing execution plans and resources in dynamic clusters using commonly-used strategies including Megatron-style 3D parallelism and ZeRO/GA/GC. It is an interesting future work to extend \ourSys{} to incorporate systems like Alpa/Unity/Whale to improve their performance in shared cluster environments.
}

\paragraph{Cluster scheduling.}
The cluster optimization for DL job has been extensively studied, with the aims of improving cluster utilization (\eg, Gandiva~\cite{gandiva2018}, AntMan~\cite{antman2020}, Lucid~\cite{lucid2023}), reducing job completion time (\eg, Tiresias~\cite{gu2019tiresias}, Optimus~\cite{optimus}),
and guaranteeing SLAs or fairness (\eg, HiveD~\cite{hived2020}, Themis~\cite{mahajan2020themis}).
Recent works~\cite{torchelastic,kungfu2020,qiao2021pollux,hwang2021coddl,li2023easyscale,gu2023elasticflow}
have further explored elasticity to optimize cluster scheduling. \revision{Sia~\cite{subramanya2023sia} is also designed for resource-adaptive jobs while supporting hybrid parallel job configurations on heterogeneous GPUs.}
However, the aforementioned studies primarily focus on data-parallel approaches, treating each individual worker as implementing a static job plan \revision{and scaling with fixed data-parallel degrees.}
This fails to align with recent trends of large language models (LLMs), which \revision{often employ various advanced parallel strategies for training.}

To achieve better DL job performance and cluster efficiency, multi-dimensional resources such as host memory~\cite{zhao2023silod}, CPUs~\cite{mohan2021analyzing, zhao2023goldminer}, and bandwidth~\cite{gandiva2018} are jointly considered for job scheduling.
Allox~\cite{allox2020} leverages the diverse resource sensitivity of DL workloads to schedule jobs between CPU and GPU resources, while
Synergy~\cite{synergy22} breaks away from proportional GPU allocation to perform resource-sensitive scheduling.
Muri~\cite{zhao2022muri} utilizes multi-resource interleaving for optimizing DL job scheduling. 
However, all these works treat DL jobs as black boxes when scheduling,
overlooking the critical opportunity presented by the varied multi-resource demands of different execution plans. 
This represents the key advantage leveraged by \ourSys{}.

\revision{QOOP~\cite{qoop} and Jiffy~\cite{anurag2022jiffy} dynamically re-evaluate and re-schedule a job’s query execution plan during its execution to improve performance for big-data computing.}
\revision{However, different from the CPU DAG execution for big-data jobs, DL training jobs are in iterative computations on GPUs. Therefore, \ourSys{} carefully considers the heterogeneous resource demands of DL training strategies for job re-planning.}

\paragraph{Performance modeling and prediction.}
DayDream~\cite{zhu2020daydream} employs fine-grained dependency graph analysis to estimate the runtime performance of DL jobs.
Habitat~\cite{yu2021habitat} leverages data gathered at runtime on one GPU to help \revision{predict another separate GPU}.
Pollux~\cite{qiao2021pollux} utilizes a model of system throughput and statistical efficiency to \revision{predict scaling performance}.
DNNPerf~\cite{gao2023runtime} leverages graph neural networks to predict GPU memory usage and iteration time.
Prior works above mainly focus on predicting performance for a single GPU or data-parallel training, while \ourSys{} takes it a step further by modeling the performance of complex parallel strategies and optimizations across multiple resources, \revision{demonstrating the great potential of improving cluster efficiency.} 

\section{Conclusion}
\label{sec:conclusion}

Looking back at the evolution of deep learning model training strategies, we find that they have always been created for 
the essential purpose of adapting the training
to different levels of resource availability, 
from a single to thousands of GPUs,
with the aid of other types of auxiliary resources. 
Such execution adaptivity and resource interchangeability are even more beneficial, but left unexplored, in shared clusters where the resources are highly dynamic.
This paper makes the first attempt at unifying the execution planning in cluster scheduling by presenting \ourSys{}, a system that shows
the great potential of this unification via: 
comprehensive performance modeling for training strategies; a multi-resource scheduling policy co-designed with execution planning; and extensive evaluations showing the vastly improved job and cluster efficiency.
We hope that \ourSys{} can inspire future advancements of both training strategies and scheduling systems, to uncover more benefits of their connection.


\bibliographystyle{ACM-Reference-Format}
\bibliography{ref}

\end{document}